\documentstyle[12pt]{article}
\input epsf
\begin{document}
\title{Quantum Teleportation with an Accelerated Observer and
Black Hole Information}
\author{
K. Shiokawa\thanks {E-mail address: kshiok@mail.ncku.edu.tw}
\\
{\small National Center for Theoretical Sciences,}\\
{\small National Cheng-Kung University, Tainan 701, Taiwan}}
\date{\today}
\maketitle
\begin{abstract}
 Nonperturbative analysis of quantum entanglement and quantum teleportation protocol
 using oscillator variables carried by observers in relativistic motion
 under the continuous influence of the environment is given.
The full time evolution of quantum entanglement among static and
accelerated observers is studied. The environment plays a dual
role. While it creates bipartite and tripartite entanglement among
observers even when the initial state is separable, it suppresses
the entanglement via decoherence. Motivated by the black hole
information problem, we consider quantum teleportation between
static and accelerated observers. Acceleration of the observer
suppresses fidelity of teleportation. Some of the quantum
information escapes outside of the horizon in the form of
bipartite and tripartite entanglement during the teleportation
process. Explicit calculation of information loss is provided. In
addition to the loss due to the interaction with the environment,
there is an intrinsic loss originated in a measurement process. We
discuss the implications of our results on the black hole case.
\end{abstract}
\newpage

\section{Introduction}

The equivalence principle tells us that a particle under constant
acceleration can be viewed as a particle under the influence of
a static gravitational field. Explicit coordinate transformation
shows that a particle moving in a hyperbolic trajectory due to
constant acceleration in Minkowski spacetime is equivalent to a
particle moving in a geodesic in Rindler spacetime. Under the
coordinate tranformation, positive frequency modes in Minkowski
spacetime are expressed as the mixture of positive and negative
frequency Rindler modes. This implies that the Minkowski vacuum
annihilated by positive Minkowski modes is not equivalent to the
Rindler vacuum. As shown in \cite{Unruh76}, an accelerated
observer sees
 the Minkowski vacuum filled with Rindler particles.
 The spectrum of Rindler particles is thermal, similar to the spectrum of the Hawking radiation from a blackhole\cite{Hawking74}.
The Unruh effect modifies the quantum dynamics of accelerated
systems in a nontrivial way causing apparent discrepancy between
the interpretation of radiation emission and absorption between
accelerated and static observers\cite{UnruhWald84}.

Stimulated by recent progress in quantum information science,
various attempts to understand relativistic quantum phenomena from
the view of quantum information have been made\cite{PeresTerno00}.
Quantum teleportation between a static and an accelerated observer
was discussed in \cite{AlsingMilburn03}, where one of the cavities
containing a single mode is assumed to follow the accelerated
trajectory. Because there was no consideration of dynamical aspects and
other crucial features in \cite{AlsingMilburn03}, their
derivation is not expected to be applicable in the general
situation\cite{SchutzholdUnruh05}. The entanglement in the system
with one accelerated and one static\cite{FuentesMann05} and both
accelerated\cite{BenattiFloreanini06,MasserSpindel06} was
discussed. Entanglement dynamics involving accelerated observers
was studied in \cite{LinHu06}.

On the problem of black hole information loss\cite{Hawking76},
entanglement in the Hawking pair is often regarded as a crucial ingredient to
rescue information from a black hole in the form of the Hawking
radiation\cite{Page93}. The nonlocal process present
in quantum teleportation is also considered to play a role\cite{DanielssonSchiffer93}.
More recently, it was speculated that
if the collapsing matter and incoming Hawking particle at the black hole
singularity\cite{MaldaneceHorowitz04} are maximally entangled,  information can
escape from a black hole as a process similar to the quantum teleportation.
 This scenario, unfortunately, is likely flawed by the interaction between the collapsing matter and the incoming Hawking flux\cite{GottesmanPreskill04},
 In the case of the qubit teleportation assuming the random unitary interaction between the original state and one of the entangled pair\cite{Lloyd06}, however,
 nearly perfect information can be restored
with proper encoding after quantum error correction.
It is of interest to see if the same scenario works in the spacetime with a horizon.
Motivated by these works, in this paper, we study the dynamical
evolution of quantum entanglement and fidelity during quantum
teleportation due to accelerated motion of observers. Near
horizon, the Rindler metric is equivalent to the Schwartzchild
metric. Thus our work can be viewed as performing a quantum
simulation of information flow around the black hole horizon.

First we develop a general framework to study the entanglement
among an arbitrary number of observers generalizing the formalism
in \cite{Shiokawa08} to the relativistic setting. Each observer's
quantum state is assumed to be described by a harmonic oscillator,
which couples linearly with a scalar quantum field. In this work,
we consider the vacuum state of the field as environment. We are
primarily interested in the quantum state of observers and trace
out all the field modes in our calculation. Then we study the
entanglement dynamics among observers' quantum states when each
observer is allowed to move in a prescribed trajectory. When one
of the observers is subjected to constant acceleration, his
worldline in a Minkowski coordinate becomes a hyperbolic curve and
asymptotically approaches his event horizon. The observer sees the
Minkowski vacuum as a thermal state. Entanglement between his and
other observers' states will be modified accordingly.

Perturbative approximations such as Born or Born-Markov
approximations are widely used for the study of decay processes
and estimation of various quantities of open systems. Radiation
processes from the accelerated particles and near black
holes\cite{BMP95} are often studied under these approximations.
However, naive perturbation expansions do not guarantee the
positivity of the reduced density matrix. Entanglement properties
are measured by the negativity of the partially transposed density
matrix and sensitive to the positivity of the density matrix. For
this reason, in this paper, we try to use an exact expression
without making perturbative approximations. Nonperturbative
analysis is one of the key achievements in solving the apparent
paradox in the radiation-reaction process viewed in inertial and
accelerated frames\cite{RSG91}.

Fluctuations of the canonical phase variables obey uncertainty
relations. Uncertainties are modified in the presence of the
environment. In the system out of equilibrium, they are
time-dependent. In order to set up the quantum teleportation
protocol\cite{BBCJPW93} using our oscillator variables, we
introduce another observer who carries a state to be teleported.
We study uncertainties in the multi-modes and follow their
dynamical evolution. We are particularly interested in the case
when one observer is under constant acceleration and study how motion of the observer affects uncertainties. Although uncertainty
relations should always be
 satisfied by any physical states in principle,
 they are no longer guaranteed under perturbative approximations.
  Similarly to entanglement,
  they are sensitive to the positivity of the density matrix.
 Next we study entanglement among these observers.
As shown in \cite{Simon}, uncertainty relations in the partially
transposed density matrix give the necessary and sufficient
condition for separability. Entanglement measures, negativity and
the log-negativity, will also be calculated from the
uncertainties.

Next we study entanglement among these three observers and clarify
the relation between bipartite and tripartite entanglement. For a
three-mode oscillator system, it is possible to find the genuine
tripartite entanglement $E_{ABC}$ from the monogamy
relation\cite{CKW00,ASI06}.
  In the conventional quantum optical setup for the teleportation\cite{BraunsteinKimble98}, the effect of vacuum is treated locally and there is no role for the tripartite entanglement. In our case, vacuum is a ground state of the quantum field which spreads nonlocally in spacetime.
   Therefore its influence on the observers will develop nonlocal correlation between their states. In order to see this, we set the initial state in a product state between a coherent and a two-mode squeezed state and see how the system develops tripartite entanglement dynamically in time.

Then we perform quantum teleportation using an entangled pair one
of which is carried by a receiver under constant acceleration. We
include the effect of quantum field continuously interacting with
the system and calculate time-dependent teleportation fidelity as
a measure of success of teleportation. First we study the case
with arbitrary final states of the sender and the carrier of one
of the entangled pair by summing over all possible states between
them. Next we consider the generalized final state to which their
state will be projected after the measurement to be in a two-mode
squeezed state. We send the measurement result to the receiver
followed by the appropriate unitary transformation on his state so
as to recover the original state with generic initial and final
squeezed states and see how much information can be recovered. We
consider both in the presence and absence of the interaction
between the system and field.

We first obtain the general expression for the fidelity for
arbitrary amount of squeezing for the initial and final state.
Then we look at the limiting case when the final state is
maximally entangled. In the conventional quantum teleportation
scheme, the final state is generally assumed to be maximally
entangled while the initial squeezing is finite due to
experimental limitation. Since our motivation is to probe black
hole information loss from the possiblly unknown final state, we
consider the generalized final state by introducing a finite
squeezing parameter. In all cases, we obtain the exact dynamical
expressions for the fidelity. We also study the asymptotic long
time limit of the fidelity and discuss the implications of our
results on the black hole information problem.

In Sec. 2, we develop general formulation for ${\cal N}$ observers
each carrying quantum state moving in a four-dimensional spacetime
interacting with quantum fields. Tracing out quantum fields, we
obtain the reduced density matrix for ${\cal N}$ observers'
quantum state. In Sec. 3, from a correlation matrix of the system,
we calculate uncertainty relations for multi-modes.
Also from the correlation matrix after partial transpose,
we calculate the bipartite entanglement measures
 for two and three modes from which we obtain genuine tripartite entanglement.
 We give a detailed study of the time evolution of these quantities under the influence of the environment in the absence and presence of acceleration of one observer.
 In Sec. 4, quantum teleportation among three parties is studied.
We first study the case by integrating out all the final states
and look at the time evolution of the teleportation fidelity. While introducing
the basic scheme of teleportation, we generalize the possible
final state to an arbitrary two-mode squeezed state. Then we
consider the effect of the environment for the protocol and study
the time evolution of the fidelity. We obtain the result for the
maximally entangled final state by taking the infinite squeezing
limit of the final state. Finally we consider the maximally
entangled initial and final states including the environment. We
will study the full time evolution and the long time, weak coupling
limit of the fidelity. We will use natural units:
$G=\hbar=c=k_B=1$.

\section{General formulation} \label{GF}

We consider the system composed of ${\cal N}$ observers moving in
prescribed trajectories. Our Hamiltonian is a relativistic
generalization of a ${\cal N}$-Brownian oscillator model studied
in \cite{Shiokawa08JMC,Shiokawa08} and given by
\begin{eqnarray}
H_S = \sum_{j=1}^{\cal N} \frac{P_j^2}{2M_j} + V_0(R_1,...,R_{\cal
N})\;, \label{HS}
\end{eqnarray}
where the bare potential $V_0$ is a sum of the physical potential
\begin{eqnarray}
V(R_1,...,R_{\cal N})= \sum_{j=1}^{\cal N}  \frac{M_j
\Omega_{j}^2}{2} R_j^2 \label{VS}
\end{eqnarray}
and the counter term $\Delta V$
\begin{eqnarray}
\Delta V(R_1,...,R_{\cal N})= \sum_{j=1}^{\cal N} \sum_{l=1}^{\cal
N} \frac{\Delta V_{jl}}{2} R_j R_l, \label{VC}
\end{eqnarray}
where
\begin{eqnarray}
\Delta V_{jl}= \frac{2 \lambda_j \lambda_l
\Lambda}{\pi}\label{VCjl}
\end{eqnarray}
with the frequency cutoff $\Lambda$ for the scalar field below.

Each oscillator variable $R_j$ couples linearly
with the scalar field $\phi_j$ at $x_j$ as
\begin{eqnarray}
H_I&=&\sum_{j=1}^{\cal N} \lambda_{j}~ R_{j}(\tau_j)
\phi_j(x_{j}(\tau_j),t(\tau_j)).\nonumber\\ \label{HI}
\end{eqnarray}
We assume that $R_j$ is an internal coordinate of a $j$-th observer which depends on
    a proper time $\tau_j$ of the observer. Our total Hamiltonian is thus Lorentz invariant.

The field Hamiltonian for each $\phi_j$ in a $D$-dimensional space is
\begin{eqnarray}
H_F&=&\frac{1}{2} \int d^D x \left[ \Pi_{\phi_j}^2 + (\partial_x
\phi_j)^2 + m_j^2 \phi_j^2 \right].
\end{eqnarray}
Each field $\phi_j$ allows a mode decomposition:
\begin{eqnarray}
  \phi_j(x,t)=
  \int \frac{d^D k}{(2\pi)^{D/2}\sqrt{2 \omega_{k}}}
\left\{ b_k(x(\tau)) e^{-i \omega_k t(\tau) + ikx(\tau)} +
b_k^{\dagger}(x(\tau)) e^{i \omega_k t(\tau)-ik x(\tau) } \right\}
 \label{field}
\end{eqnarray}
with $ \omega_k \equiv \sqrt{k^2 + m^2 }$.

 The Heisenberg equations that $R_j$ satisfy are
\begin{eqnarray}
M_j \ddot{R}_j(\tau) + M_j \Omega_{j}^2 R_j + \sum_{l} \Delta V_{jl}
R_j + 2 \sum_{l} \int_{0}^{\tau_j} ds \alpha_{Ijl}(\tau_j,s_l) R_l(s_l)=0,
\label{EL1}
\end{eqnarray}
where
\begin{eqnarray}
\alpha_{Ijl}(\tau_j,\tau'_l)= - \lambda_{j}^2 \sum_k \cos \left[ k
(x_j(\tau_j)-x_l(\tau'_l)) \right] \sin \left[ \omega_k
(t(\tau_j)-t(\tau'_l))\right]/2\omega_k \label{alphaI}
\end{eqnarray}
is an imaginary part of the response function\cite{FeyVer63} defined
as \\$\alpha_{jl}(\tau_j,\tau'_l)\equiv \lambda_{j}^2 \sum_k \cos \left[
k (x_j(\tau_j)-x_l(\tau'_l)) \right]
 e^{-i\omega_k
(t(\tau_j)-t(\tau'_l))}/2\omega_k$, where $\sum_k \equiv \int d^D
k/(2\pi)^D$. We set the initial condition at $t=0$ hypersurface
when all observers are at rest with their proper times set to
$\tau_j=0$ for all $j$. Eq.(\ref{EL1}) contains nonlocal kernels
$\alpha_{Ijl}(\tau_j,\tau'_l)$ and the exact time evolution of
$R_j$ at $\tau$ depends on its and other observer's past history.
Thus the dynamics of $R_j$ is not Markovian.

\begin{figure}[h]
 \begin{center}
\epsfxsize=.45\textwidth \epsfbox{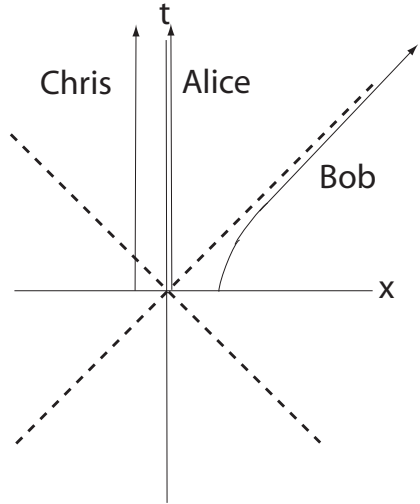}
\end{center}
\caption{ The world line of three observers.\label{fig1}}
\end{figure}
 Let us consider the case ${\cal N}=3$ in which the first observer Bob is accelerated with a constant acceleration $a$ in an accelerated trajectory so
that its coordinate is given by $t(\tau)=\sinh(a\tau)/a$ and
$x_1(\tau)=\cosh(a\tau)/a$. The second observer Alice and the third observer Chris's locations are fixed at $x_2=0$ and $x_3=-L <0$, respectively.
We assume that Alice and Bob are close to each other ($\epsilon<<1$)
and share a common environment ($\phi_2=\phi_3$), while the
accelerated observer Bob is located in a distance such that $R_1$ can
be treated as independently coupling to an environment. The response functions
(\ref{alphaI}) are then given by
\begin{eqnarray}
\alpha_{11}(\tau,\tau')&=& \lambda_{1}^2 \sum_k \cos \left[ k
(x_1(\tau)-x_1(\tau')) \right]
 e^{-i\omega_k
(t(\tau)-t(\tau'))}/2\omega_k, \nonumber\\
\alpha_{jj}(\tau,\tau')&=& \lambda_{j}^2 \sum_k
 e^{-i\omega_k
(\tau-\tau')}/2\omega_k  ~(\mbox{for} ~j=2,3), \nonumber \\
\alpha_{23}(\tau,\tau')&=&\alpha_{32}(\tau,\tau')=
\lambda_{2}\lambda_{3} \sum_k
 \cos (L k) e^{-i\omega_k
(\tau-\tau')}/2\omega_k,   \label{alphaI2}
\end{eqnarray}
and $\alpha_{12}=\alpha_{13}=\alpha_{21}=\alpha_{31}=0$.
Note that the coordinate time $t$ and the proper time  are the same for Alice and Chris but not for Bob.
The most interesting term above is $\alpha_{11}(\tau)$ where each mode
in Minkowski spacetime is expressed in Rindler coordinates.
The conversion of modes in different spacetimes is given by the Bogoliubov transformation between annihilation/creation operators\cite{BirrelDavis}.
Writing the momentum space integral in (\ref{field}) in terms of Rindler modes reproduces a standard derivation of the Unruh effect based on the Bogoliubov transformation\cite{BirrelDavis}.
For a massless field $\phi_{1,2}$ in a three dimensional space, the kernels in (\ref{alphaI2}) can be brought to the following form:
\begin{eqnarray}
\alpha_{11}(\tau)&=& \frac{\lambda_{1}^2}{2\pi^2} \int_{0}^{\Lambda}
d k k \left[  \cos(k \tau) \coth(\frac{\pi k}{a})-i\sin (k \tau)
\right], \nonumber\\
 \alpha_{jj}(\tau)&=& \frac{\lambda_{j}^2}{2\pi^2}
\int_{0}^{\Lambda} dk k \left[  \cos(k \tau)-i\sin (k \tau) \right],
~(\mbox{for} ~j=2,3), \nonumber\\
 \alpha_{23}(\tau)&=& \alpha_{32}(\tau) = \frac{\lambda_{2}\lambda_{3}}{\pi^2 L}
\int_{0}^{\Lambda} dk k \sin (L k) \left[ \cos(k \tau)-i\sin (k
\tau) \right].
\end{eqnarray}
The kernel $\alpha_{23}$ is responsible for the interaction between
Alice and Chris's state that becomes negligible for large
separation as $1/L$, which is a characteristic decay of a  massless field in three dimension and justifies neglecting the interaction between the accelerated
observer and the rest of observers.
Comparing the first kernel $\alpha_{11}$ with that of the influence
functional\cite{FeyVer63}, we see that the first observer sees the
vacuum as a thermal bath with the Unruh temperature
$T_U=a/2\pi$\cite{Anglin93,RHA96}.
   Writing the influence kernel directly in the Rindler time has advantages in studying quantum dynamical aspects of observer's quantum states under the influence of the Unruh effect, which we will describe below in details.

For large $\Lambda$ and small separation $L$,  the Heisenberg
equations of motion for three coordinate variables can be written in the
local form:
\begin{eqnarray}
M_1 \ddot{R}_1(t) &+& M_1 \Omega_{1}^2 R_1(t) + \gamma_1
\dot{R}_1(t)=0, \label{ELNR1}\\
M_2 \ddot{R}_2(t) &+& M_2 \Omega_{2}^2 R_2(t) + \gamma_2
\dot{R}_2(t)  + \gamma \dot{R}_3(t)=0, \label{ELNR2}\\
M_3 \ddot{R}_3(t) &+& M_3 \Omega_{3}^2 R_3(t) + \gamma_3
\dot{R}_3(t)  + \gamma \dot{R}_2(t)=0,\nonumber
\end{eqnarray}
where we write $\gamma_j\equiv \lambda_j^2/\pi $ (for $j=1,2,3$) and
$\gamma \equiv \lambda_2 \lambda_3/\pi$. The divergence from
nonlocal terms are canceled by counter terms. For our case,
$\Delta V_{12}=\Delta V_{21}=\Delta V_{13}=\Delta V_{31}=0$. Hereafter we will set $M_1=M_2=M_3=1$.

We write the solution for the first equation in (\ref{ELNR1}) with
initial conditions $R_1(0)=0$ and $\dot{R}_1(0)=1$ as $g_1(\tau)$:
\begin{eqnarray}
g_{1}(\tau) = \frac{\sin(\Omega_{1r}
\tau)}{\Omega_{1r}}e^{-\gamma_1\tau},
\label{solutiong1}
\end{eqnarray}
where $\Omega_{1r}^2 \equiv \Omega_1^2 - \gamma_1^2$.
A pair of solutions of (\ref{ELNR2}) with initial
conditions $R_2(0)=R_3(0)=0$ and $\dot{R}_2(0)=1$, $\dot{R}_3(0)=0$
will be written as $h_3(\tau)$ and $h_5(\tau)$, respectively. For
$\Omega_{2}=\Omega_{3}$ and $\lambda_2=\lambda_3$, the solutions are
given by $h_3(t)\equiv (g_{3}(t)+g_{0}(t))/2$ and $h_5(t)\equiv
(g_{3}(t)-g_{0}(t))/2$, where
\begin{eqnarray}
g_{3}(t) = \frac{\sin(\Omega_{2r} t)}{\Omega_{2r}}e^{-\gamma t}
~\mbox{and}~ g_{0}(t) = \frac{\sin(\Omega_2 t)}{\Omega_2}
\label{solution2}
\end{eqnarray}
are the solutions corresponding to two normal modes of a coupled
oscillator and $\Omega_{2r}^2 \equiv \Omega_2^2 - \gamma^2$.

General solutions with arbitrary initial conditions $R_{j0}$ and
$P_{j0}$ of the Heisenberg equations (\ref{ELNR1}) and
(\ref{ELNR2}) for $j=1,2,3$ are
\begin{eqnarray}
R_j(\tau) &=&
\sum_{k=1}^{3} C_{R_{j}R_{k}} R_{k0} +  \sum_{k=1}^{3} C_{R_{j}P_{k}} P_{k0}
   +\lambda_j \int_{0}^{\tau} ds g_{2l_j-1}(\tau-s) {\phi}_j(s),
\nonumber\\ P_j(\tau)  &=&
\sum_{k=1}^{3} C_{P_{j}R_{k}} R_{k0} +  \sum_{k=1}^{3} C_{P_{j}P_{k}} P_{k0}
   +\lambda_j \int_{0}^{\tau} ds g_{2l_j}(\tau-s) {\phi}_j(s),
\label{ClassicalEvolution}
\end{eqnarray}
where ${g}_{2j}\equiv\dot{g}_{2j-1}$ and $l_1=1,l_{2,3}=2$.

The expectation value of phase space variables can be expressed in the
matrix form:
\begin{eqnarray}
\left( \begin{array}{c}
        \langle R_1 \rangle \\
        \langle P_1 \rangle \\
        \langle R_2 \rangle \\
        \langle P_2 \rangle \\
        \langle R_3 \rangle \\
        \langle P_3 \rangle
         \end{array}      \right)
         =
      {\cal C}
      \left( \begin{array}{c}
          R_{10} \\
          P_{10} \\
          R_{20} \\
          P_{20} \\
          R_{30} \\
          P_{30}
         \end{array}      \right)
=  \left( \begin{array}{cccccc}
      C_{R_1 R_1}&C_{R_1 P_1}
     &0&0&0&0\\
       C_{P_1 R_1}&C_{P_1 P_1}
     &0&0&0&0\\
       0&0&C_{R_2 R_2}&C_{R_2 P_2}&C_{R_2 R_3}&C_{R_2 P_3}\\
       0&0&C_{P_2 R_2}&C_{P_2 P_2}&C_{P_2 R_3}&C_{P_2 P_3}\\
       0&0&C_{R_3 R_2}&C_{R_3 P_2}&C_{R_3 R_3}&C_{R_3 P_3}\\
       0&0&C_{P_3 R_2}&C_{P_3 P_2}&C_{P_3 R_3}&C_{P_3 P_3}
            \end{array}      \right)
     \left( \begin{array}{c}
          R_{10} \\
          P_{10} \\
          R_{20} \\
          P_{20} \\
          R_{30} \\
          P_{30}
         \end{array}      \right)
\label{RC2}
\end{eqnarray}
The time evolution matrix $ {\cal C}$ for our solutions is given by
\begin{eqnarray}
  {\cal C}
 \equiv
  \left( \begin{array}{cccccc}
   f_1 & g_1 & 0 & 0 & 0& 0\\
   f_2 & g_2 & 0 & 0 & 0& 0\\
   0 & 0 & f_3 & h_3 & f_5 & h_5\\
   0 & 0 & f_4 & h_4 & f_6 & h_6\\
   0 & 0 & f_5 & h_5 & f_3 & h_3\\
   0 & 0 & f_6 & h_6 & f_4 & h_4\\
   \end{array}      \right),
     \label{C}
\end{eqnarray}
where $f_{2j}\equiv \dot{f}_{2j-1}$, and $f_{2j-1}\equiv
{g}_{2j}-2\dot{g}_{2j}(0) g_{2{l}_j-1}$ for $j=1$ and
$f_{2j-1}\equiv {h}_{2j}-2\dot{h}_{2j}(0) g_{2{l}_j-1}$ for
$j=2,3$.

It is convenient for our purpose to use a characteristic function
for the Wigner representation\cite{Wigner32}. For the
${\cal N}$-particle system:
\begin{eqnarray}
  \chi_{W}({\cal Y},\tau)
  &=& \mbox{Tr} \left[ {\rho}(0)
  e^{i \sum_{j=1}^{2\cal N} Y_j {X}_j(\tau)}
            \right],
   \label{chiW}
    \end{eqnarray}
   where we defined
   ${X}_{2j-1}\equiv \sqrt{\Omega_j} {R}_j$,
   ${X}_{2j}\equiv  {P}_j/\sqrt{\Omega_j}$
   for $j=1,2,...,{\cal N}$
   and ${\cal Y}\equiv(Y_1 ... Y_{2\cal N})$. The matrix ${\cal C}$
   is scaled accordingly.
Now we trace out the field $\phi$ in order to obtain the reduced
dynamics of the system. In our case ${\cal N}=3$ with a factorized
initial condition: ${\rho}(0) =
  {\rho}_{S}(0) \otimes \prod_{j=1}^{2} {\rho}_{\phi_j}(0)$,
  $\chi_{W}$ is also factorized into two parts as
  $ \chi_{W}({\cal Y},\tau) = \chi_{W}^S({\cal Y},\tau)
  \prod_{j=1}^{2} \chi_{W}^{\phi_j}({\cal Y},\tau)$.
The system part $\chi_{W}^S({\cal Y},\tau)=\mbox{Tr}_{S} \left[
{\rho}_S(0) e^{i \sum_{j=1}^{2\cal N} Y_j {X}_{Cj}(\tau)}
            \right] $, where ${X}_{Cj}(\tau)$ are solutions of the Heisenberg
            equations with $\phi_j=0$, and the environment parts
\begin{eqnarray}
\chi_{W}^{\phi_1}({\cal Y},\tau)&=& \mbox{Tr}_{\phi_1} \left[
{\rho}_{\phi_1}(0)
  \exp\left[ i \lambda_1 \sum_{l=1}^{2}  \left\{ Y_l
 \int_{0}^{\tau} ds
g_{l}(\tau-s) \right\} {\phi_1}(s) \right] \right],\nonumber\\
\chi_{W}^{\phi_2}({\cal Y},\tau)&=& \mbox{Tr}_{\phi_2} \left[
{\rho}_{\phi_2}(0)
  \exp\left[ i \left\{ \sum_{l=3}^{6} \lambda_{[l]} Y_l
 \int_{0}^{\tau} ds
g_{(l)}(\tau-s)  \right\} {\phi_2}(s) \right] \right],
            \label{chiB}
\end{eqnarray}
where $[l]$ is equal  to $l/2$ (for even $l$) and $(l+1)/2$ (for odd
$l$) and $(l)=l$ for $l=3,4$ and $(l)=l-2$ for $l=5,6$.
In response to the state of the system, each field mode is shifted
by system-bath interaction.
After the time scale of each field mode, the back action of this process changes
the system's state depending on the shift of the field mode.
   The collective effect of these processes
   leads to the non-Markovian evolution of the reduced system.

The field characteristic function in (\ref{chiB}) can be evaluated
exactly. We assume the environment is initially in a Minkowski
vacuum state. Its density matrix is given as ${\rho}_{\phi}(0)=
\mid 0_M \rangle \langle 0_M \mid$.  The environment
characteristic function can be written as
\begin{eqnarray}
\chi_{W}^{\phi_1}({\cal Y},t) \chi_{W}^{\phi_2}({\cal Y},t) &=&
\exp\left[ -\frac{1}{2} {\cal Y}^T
      {\bf \Large \Sigma} {\cal Y}
\right]\\ &=& \exp\left[ -\frac{1}{2}(Y_1 ... Y_{6})^T
\left( \begin{array}{cccccc}
      \Sigma_{11} & \Sigma_{12}
     &0&0&0&0\\
       \Sigma_{21}&\Sigma_{22}
     &0&0&0&0\\
       0&0&\Sigma_{33}&\Sigma_{34}&\Sigma_{35}&\Sigma_{36}\\
       0&0&\Sigma_{43}&\Sigma_{44}&\Sigma_{45}&\Sigma_{46}\\
       0&0&\Sigma_{53}&\Sigma_{54}&\Sigma_{55}&\Sigma_{56}\\
       0&0&\Sigma_{63}&\Sigma_{64}&\Sigma_{65}&\Sigma_{66}
            \end{array}      \right)
        \left(   \begin{array}{c}
        Y_1 \\ ...\\ Y_{6} \\
         \end{array}      \right)
\right], \nonumber
   \label{chiBGauss}
\end{eqnarray}
where
\begin{eqnarray}
\Sigma_{jl}(\tau) &=& \left\{ \begin{array}{c}
\frac{\lambda_1^2}
  { 4 \pi } \int_{-\infty}^{\infty} \frac{dk}{\omega_k}
 \int_{0}^{\tau} ds \int_{0}^{\tau} ds' g_{j}(\tau-s)
g_{l}(\tau-s')
e^{i k (x_1(s)-x_1(s'))} e^{-i\omega_k (t_1(s)-t_1(s'))}
~\mbox{for} ~j,l=1,2 \\
  \frac{\lambda_{[j]}\lambda_{[l]}}
  { 2 \pi } \int_{0}^{\infty} \frac{d\omega}{\omega}
   \int_{0}^{\tau} ds \int_{0}^{\tau} ds' g_{(j)}(\tau-s) \cos
\omega (s-s') g_{(l)}(\tau-s')
 ~\mbox{for} ~j,l=3,...,6
 \end{array} \right.
\nonumber
   \label{sigma}
\end{eqnarray}
are time-dependent, nonequilibrium fluctuations of the system
variables induced from the environment.
Nonvanishing off-diagonal correlations $\Sigma_{jl}$ for $j\neq l$
generated from the interaction with the environment are signatures of induced correlation and
entanglement among observer's quantum states.
For $j,l=1,2$,
\begin{eqnarray}
\Sigma_{jl}(\tau) &=&
 \frac{\lambda_1^2}
  { 2 \pi } \int_{0}^{\infty} d\omega \omega \coth(
\frac{\omega \pi}{a})
 \int_{0}^{\tau} ds \int_{0}^{\tau} ds' g_{j}(\tau-s)
g_{l}(\tau-s')
\cos \omega (s-s').
   \label{sigma34}
\end{eqnarray}

If the initial states of system variables are all Gaussian states
with vanishing mean positions and
momenta, $\langle{\cal X}(0)\rangle=0$, the system characteristic
function also takes the Gaussian form:
\begin{eqnarray}
\chi_{W}^S({\cal Y},t)&=& \exp\left[ -\frac{1}{2}{\cal Y}^T
   (\Delta {\cal X})_{C}^2(\tau) {\cal Y}
  \right] \\
  &\equiv& \exp\left[ -\frac{1}{2}(Y_1 ... Y_{6})^T
  \left( \begin{array}{ccc}
\langle  \{X_{1C}, X_{1C}\}  \rangle &...&\langle \{
X_{C1}, X_{C6}\}\rangle \\& ... &\\
\langle \{X_{C6}, X_{C1}\}\rangle
   &...&\langle  \{X_{C6}, X_{C6}\}\rangle
    \end{array}\right)
       \left(   \begin{array}{c}
        Y_1 \\ ...\\ Y_{6} \\
         \end{array}      \right)
\right], \nonumber
   \label{chiBS}
\end{eqnarray}
where $ \{A,B\} \equiv (AB+BA) /2$ is an anticommutator and ${\cal
X}_{C}=(X_{C1} ...X_{C6})$ satisfy the equations of motion
(\ref{ELNR1}) and (\ref{ELNR2}) for damped harmonic oscillators.
$(\Delta {\cal X})_{C}^2(\tau)$ are essentially the initial
fluctuations of the system variables shifted by the damped
oscillatory motion of a coupled harmonic oscillator. Combining
with the characteristic function for the field, we obtain
\begin{eqnarray}
 \nonumber\\
& &\chi_{W}({\cal Y},t) = \exp\left[ -\frac{1}{2}\cal{Y}^T
 \langle \{\cal{X},\cal{X}^{T}\} \rangle \cal{Y}
  \right] \\
&=& \exp\left[ -\frac{1}{2}(Y_1 ... Y_{6})^T
  \left( \begin{array}{ccc}
    \langle  \{X_{1}, X_{1}\}  \rangle & ... &
    \langle  \{X_{1}, X_{6}\}\rangle
      \\
       & ... &  \\
     \langle  \{X_{6}, X_{1}\}\rangle  & ... &
       \langle  \{X_{6}, X_{6}\}\rangle
            \end{array}      \right)
        \left(   \begin{array}{c}
        Y_1 \\ ...\\ Y_{6} \\
         \end{array}      \right)
\right]  \nonumber\\ &=& \exp\left[ -\frac{1}{2}(Y_1 ... Y_{6})^T
  \left( \begin{array}{ccc}
\langle  \{X_{1C}, X_{1C}\}  \rangle +\Sigma_{11}&...&\langle \{
X_{C1}, X_{C6}\}\rangle +\Sigma_{16}\\& ... &\\
\langle \{X_{C6}, X_{C1}\}\rangle
     +\Sigma_{61}&...&\langle  \{X_{C6}, X_{C6}\}\rangle
       + \Sigma_{66}     \end{array}\right)
       \left(   \begin{array}{c}
        Y_1 \\ ...\\ Y_{6} \\
         \end{array}      \right)
\right]. \nonumber
   \label{chiBW}
\end{eqnarray}

\section{Entanglement dynamics of accelerated oscillators}\label{2ENTACC}
A separability criterion for a bipartite two-level-system can be
naturally extended to continuous Gaussian variables\cite{Simon}.
The necessary and sufficient condition for separability of the
density matrix is to have only non-negative eigenvalues after the
partial transpose of one of its subsystem. For Gaussian variables
in the Wigner distribution, the partial transpose of a density
matrix in one of the oscillator component is equivalent to a
mirror reflection of that component. The necessary and sufficient
condition for the general $(1+{\cal N})$ bipartite Gaussian modes
to be separable is that the partially mirror reflected state is
still a physical quantum state that satisfies the uncertainty
principle\cite{Serafini06}. For instance, for three oscillators,
the bipartite entanglement between the first and the rest of
oscillators can be measured by first taking a partial mirror
reflection on the first variable in the phase space as
$(R_1,P_1,R_2,P_2,R_3,P_3)\rightarrow(R_1,-P_1,R_2,P_2,R_3,P_3)$,
then seeing if the resulting state satisfies or violates the
uncertainty relation. Note that the total mirror reflection does
not change the entanglement properties thus the reflection on the
second and third variables yields the same result. In terms of
${\cal X}$, this can be expressed as a matrix multiplication by
the matrix $\eta\equiv \mbox{diag}(1,-1,1,1,1,1)$ as ${\cal
X}\rightarrow \eta {\cal X}$. It follows that the partial mirror
reflection transforms the covariance matrix as
\begin{eqnarray}
(\Delta {\cal X})^2 \rightarrow \eta (\Delta {\cal X})^2
\eta^T.\label{XbyPT}\end{eqnarray}

From Williamson's theorem\cite{Williamson36}, there exists a
symplectic transformation that diagonalizes any positive-definite
$2n \times 2n$ symmetric matrix into the following form:
\begin{eqnarray}
  (\Delta {\cal X}_D)^2 = \left( \begin{array}{cccc}
   \zeta_1 & 0       & ... & 0  \\
      0    & \zeta_1 & ... & 0 \\
      ...  & ... & ... & ... \\
      0    & ... & \zeta_n & 0     \\
      0    & ... & 0     & \zeta_n     \end{array}      \right).
   \label{Williamson}
\end{eqnarray}
Although such a symplectic transformation does not preserve the
eigenvalue spectrum in general, the diagonal components $\zeta_l$
for $l=1...{\cal N}$ can be calculated as follows. Writing a commutation
relation in a $2{\cal N}\times 2{\cal N}$ matrix form as
$\left[X_i,X_j\right]=i\Gamma_{ij}$ with
\begin{eqnarray}
 \Gamma =
\left( \begin{array}{cccc}
      0    & 1 & ...   & 0  \\
      -1   & 0 & ...   & 0  \\
      ...  & ... & ... & ... \\
      0    & ... & 0   & 1     \\
      0    & ... & -1  & 0    \end{array}      \right),\nonumber
   \label{Gamma}
\end{eqnarray}
we construct a real symmetric matrix $\Delta {\cal X} \Gamma
(\Delta {\cal X})^2 \Gamma^{T} \Delta {\cal X}$. This matrix has
an eigenvalue spectrum $\zeta_l^2$ ($l=1...{\cal
N}$)\cite{SMD:94}.
The uncertainty relation can be generalized to
a symplectic invariant form $(\Delta {\cal X})^2+i\Gamma/2\geq0$.
By changing to the diagonalized form $(\Delta {\cal X}_D)^2$
above, the uncertainty relation is equivalent to saying that
$\zeta_l \geq 1/2$ for all $l$.

\subsection{Bipartite entanglement}
For a bipartite continuous variable system, there is a criteria
for a necessary and sufficient condition for
separability\cite{Simon,Peres,Horodecki97}. We write the $2\times
2$ covariance matrix here as
\begin{eqnarray}
(\Delta {\cal X})^2 \equiv \langle \{\cal{X},\cal{X}^{T}\} \rangle
\equiv \left(
\begin{array}{cc}
   D_1   & A    \\
   A^{T} & D_2  \end{array}      \right),
    \label{V2}
\end{eqnarray}
then eigenvalues $\zeta_l^2$ can be written explicitly in terms of
the following symplectic invariants $\Delta_{1,2}$ constructed
from the determinants of covariances $|A|$, $|D_{1,2}|$, $|(\Delta
{\cal X})^2|$ as
\begin{eqnarray}
\begin{array}{lll}
\Delta_1 &=& |D_1|+|D_2|+2|A|,\\ \nonumber
\Delta_2 &=& |(\Delta {\cal X})^2|.
\end{array}
  \label{Sympinv}
\end{eqnarray}
Then
\begin{eqnarray}
\begin{array}{c}
\zeta_{\pm}^2 = \frac{1}{2}\left[ \Delta_1 \pm \sqrt{\Delta_1^2-4\Delta_2 }\right].
\end{array}
  \label{zetaSym}
\end{eqnarray}
Under the partial transpose (\ref{XbyPT}) without the third varible, $A \rightarrow -A$ and
these eigenvalues will be changed to
\begin{eqnarray}
\begin{array}{c}
\lambda_{\pm}^2 = \frac{1}{2}\left[ \Delta_1 \pm \sqrt{\tilde{\Delta}_1^2-4\Delta_2 }\right],
\end{array}
  \label{lambdaSym}
  \end{eqnarray}
  where $\tilde{\Delta}_1\equiv |D_1|+|D_2|-2|A|$.
The separability conditions are
\begin{eqnarray}
\begin{array}{c}
\lambda_{\pm}^2 \geq \frac{1}{4}.
\end{array}
  \label{SEPinv}
  \end{eqnarray}
Note that the inequalities for $\zeta_{+}$ and $\lambda_{+}$ follow automatically from those for $\zeta_{-}$ and $\lambda_{-}$. Thus $\lambda_{-}$ carries the
essential information on the separability of quantum states.

Let us consider a two-mode squeezed state with a squeezing
parameter $r$ as an initial state\cite{QOtext}. Its correlation
matrix is
\begin{eqnarray}
(\Delta {\cal X})_{C}^2(0)\equiv \langle  \{{\cal X}_{C}(0),{\cal
X}^{T}_{C}(0) \}  \rangle &=&
\frac{1}{2}  \left( \begin{array}{cc} \cosh(2r) {\bf  1} & -\sinh(2r)\sigma_3 \\
-\sinh(2r) \sigma_3 & \cosh(2r) {\bf 1}
                \end{array}      \right).\nonumber
               \label{Cmatrix}
\end{eqnarray}
In the Wigner representation, the same state can be expressed as
\begin{eqnarray}
W(R_1,R_2,P_1,P_2)=\frac{4}{\pi^2} e^{-e^{2r}\left[ \Omega
(R_1-R_2)^2+(P_1+P_2)^2/\Omega\right] -e^{-2r}\left[
\Omega(R_1+R_2)^2+(P_1-P_2)^2/\Omega\right]}.
               \label{WEPR}
\end{eqnarray}  This state can be
obtained by acting a squeezing operator $e^{ir(R_1P_2-P_1R_2) }$
on the vacuum. For a large squeezing $r\rightarrow \infty$,
$W(R_1,R_2,P_1,P_2)\sim \delta(R_1-R_2)\delta(P_1+P_2)$, thus it becomes the
EPR (Einstein-Podolsky-Rosen) state\cite{EPR:35}.

In Fig. 2, the temporal behavior of the uncertainty
$\zeta_{-}$ as a function of the proper time of Bob is plotted. The initial state is a pure two mode
squeezed state introduced above. This state satisfies the uncertainty
relation with the minimum uncertainty $1/4$. As the state becomes
mixed, the uncertainty first increases in time and oscillates periodically.
Acceleration of the observer yields larger uncertainty by making the amplitude of oscillations larger.
\begin{figure}[h]
 \begin{center}
\epsfxsize=0.8\textwidth \epsfbox{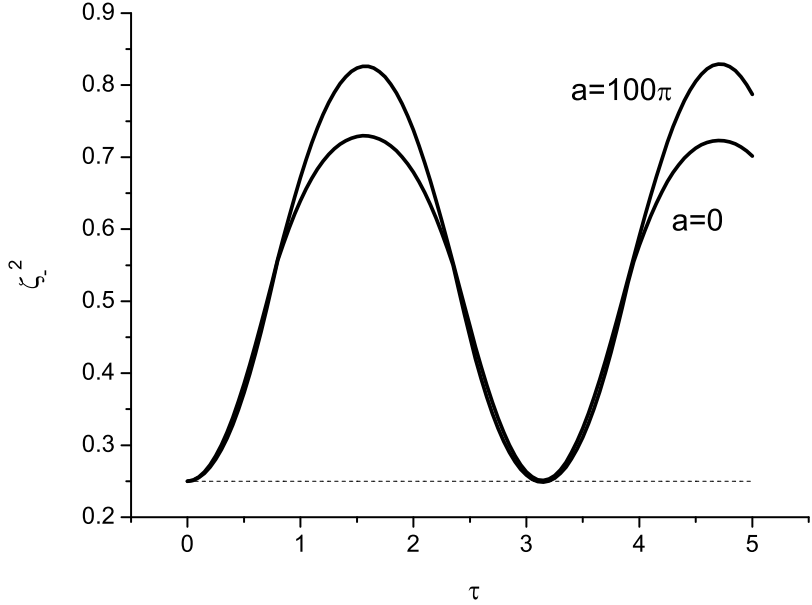}
\end{center}
\caption{ Uncertainties
 ( $\zeta_{-}$ in Eq. (\ref{zetaSym}) ) in the presence and absence of Bob's acceleration are plotted.
 The initial condition is a two-mode squeezed state with $r=0.1$. $\Omega$ is set to $1$ for all numerical plots. Other
 parameters are
$\gamma=0.001$,~$\Lambda=100$. \label{fig2}}
\end{figure}
In Fig. 3, the temporal behavior of $\zeta_{-}$ (uncertainty)  and $\lambda_{-}$ (uncertainty after the partial transpose)  is plotted. The initial state is a pure two mode
squeezed state with $r=0.5$.
The state exhibits environment-induced disentanglement at $\Omega t\sim 0.4$ with no acceleration and at
$\Omega t\sim 0.31$ with acceleration. The role of acceleration is similar to that of temperature that increases the disentanglement rate.

Now we study entanglement measures; the negativity and the log negativity.
The negativity $\cal{N}$\cite{VidalWerner02,ASI04} for our system can be defined as
\begin{eqnarray}
\cal{N}&=&\frac{||\rho_r^{T}||-1}{2},
  \label{negativity}
   \end{eqnarray}
where $\rho_r^{T}$ is the reduced density matrix after the partial
transpose. $\cal{N}$ is equal to the sum of all negative
eigenvalues of $\rho_r^{T}$. Thus it measures how much $\rho_r^{T}$
fails to be positive. The Peres criteria\cite{Peres} tell us that this can be used as a measure of entanglement. It is an entanglement monotone, which does not increase under local operations and classical communications. The logarithmic
negativity $E_{\cal{N}}$ defined as
\begin{eqnarray}
E_{\cal{N}}&=&\ln ||\rho_r^{T}||
  \label{lognegativity}
   \end{eqnarray}
  is also an entanglement monotone.
The diagonalization of $(\Delta {\cal X})^2$ changes the
   original state into the thermal state. Then the partially
   transposed density matrix $\rho_r^{T}$ after the same transformation
   also has the thermal form.
 It can be written in terms of
   the symplectic invariants $\lambda_{\pm}$ as
   \begin{eqnarray}
\rho_r^{T} &=& \prod_{\pm} \left[ \left(\frac{2}{2\lambda_{\pm}+1}
\right)
 \sum_{n=0}^{\infty} \left(\frac{2\lambda_{\pm}-1}{2\lambda_{\pm}+1}
 \right)^n
|n_{\pm} \rangle \langle n_{\pm} | \right].
  \label{rhoTthermal}
   \end{eqnarray}
   For separable states, $\lambda_{\pm}\geq
   1/2$. Then $||\rho_r^{T}||=1$ and ${\cal N}=E_{\cal N}=0$.
   For entangled states, $\lambda_{-}<1/2$ but $\lambda_{+}\geq
   1/2$.
   Thus both $\cal{N}$ and
$E_{\cal{N}}$ can be expressed in terms of $\lambda_{-}$ as
\begin{eqnarray}
\cal{N}&=&\mbox{max}
\left[0,\frac{1-2\lambda_{-}}{4\lambda_{-}}\right], \nonumber\\
E_{\cal{N}}&=&\mbox{max}\left[0, -\ln (2 \lambda_{-})\right].
  \label{negativity2}
   \end{eqnarray}
 In Fig. 4, the negativity ${\cal{N}}$ and the logarithmic negativity $E_{\cal{N}}$ are shown as a
function of the proper time. The initial state is a two mode squeezed state. They both
vanish at the same time since they both give the necessary and sufficient condition of bipartite entanglement. The larger acceleration yields the larger disentanglement rate.

\begin{figure}[h]
 \begin{center}
\epsfxsize=0.7\textwidth \epsfbox{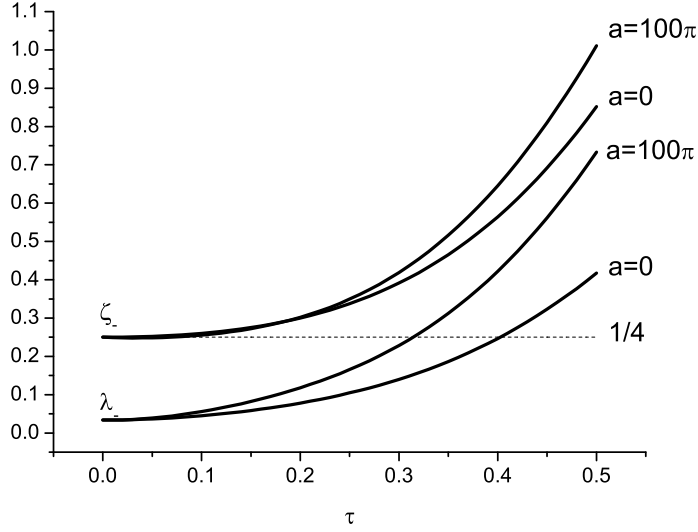}
\end{center}
\caption{ The temporal evolution of uncertainties  (
$\zeta_{-}$ in Eq. (\ref{zetaSym}) and $\lambda_{-}$ in Eq.
(\ref{lambdaSym}) ) before and after the partial transpose in the presence and absence of Bob's acceleration is plotted.
 The initial condition is a two mode squeezed state with $r=0.5$.  Other
 parameters are $\gamma=0.05$,~$\Lambda=50$.
\label{fig3}}
\end{figure}
\begin{figure}[h]
 \begin{center}
\epsfxsize=0.7\textwidth \epsfbox{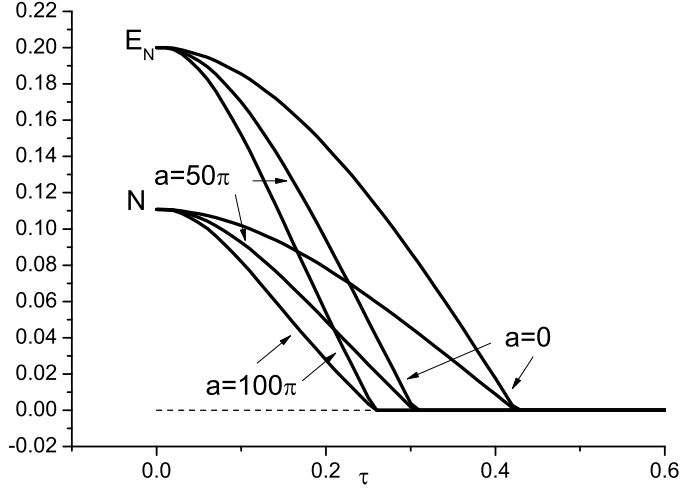}
\end{center}
\caption{ The temporal evolution of the negativity (\ref{negativity})  and
the lognegativity (\ref{lognegativity}) for different acceleration before and
after the partial transpose is plotted.
 The initial condition is a two mode squeezed state with $r=0.1$.  Other
 parameters are
$\gamma=0.1$,~$\Lambda=50$.
\label{fig4}}
\end{figure}

\subsection{Tripartite entanglement dynamics} \label{3ENT}

It is convenient to characterize an $n$-mode uncertainty relation as a sum of
symplectic invariants\cite{Serafini06} as
\begin{eqnarray}
\Sigma_n\equiv
\sum_{j=0}^{n}4^{j-n}(-1)^{n+j}\Delta_{j}^{n},
  \label{Sigma}
   \end{eqnarray}
where $\Delta_{j}^{n} (j=1,...,n)$
are principal minors\cite{Jeffreys99} of the matrix $\Gamma (\Delta {\cal X})^2$ of order $2j$.
We defined $\Delta_{0}^{n}\equiv 1$.

For the matrix in the Williamson normal form (\ref{Williamson}), $\Delta_{j}^{n} =\sum_{\left\{k_1,...,k_j\right\}}\zeta_{k_1}^2...\zeta_{k_j}^2 (j=1,...,n)$,
  where $\left\{k_1,...,k_j\right\}$ is an unordered set of different integers between $1$ and $n$.
  Then we have
\begin{eqnarray}
\Sigma_n&=& \frac{(-1)^{n}}{4^n} (
1 - \sum_{k_1=1}^{n}4\zeta_{k_1}^2+...
+(-4)^{n}\sum_{k_1=...=k_n=1}^{n}\zeta_{k_1}^2...\zeta_{k_n}^2)
\\ \nonumber
        &=& (\zeta_{1}^2-\frac{1}{4})...(\zeta_{n}^2-\frac{1}{4}).
  \label{Sigmanormal}
   \end{eqnarray}
Thus the uncertainty relation implies $\Sigma_n\geq 0$.

For the tripartite case $n=3$, we write
\begin{eqnarray}
(\Delta {\cal X})^2 =\left(
\begin{array}{ccc}
   D_1        &  A_{12}     & A_{13} \\
   A^{T}_{12} &  D_2        & A_{23}     \\
   A^{T}_{13} &  A^{T}_{23} & D_3
          \end{array}      \right).
    \label{V23}
\end{eqnarray}
Then
\begin{eqnarray}
\begin{array}{lll}
\Delta_{1}^{3} &=& |D_1|+|D_2|+|D_3|+2|A_{12}|+2|A_{13}
|+2|A_{23}|,\\
\Delta_{2}^{3} &=& |\hat{D}_1|+|\hat{D}_2|+|\hat{D}_3|+2|\hat{A}_{12}|+2|\hat{A}_{13}|+2|\hat{A}_{23}|,
\\ \Delta_{3}^{3} &=& |(\Delta {\cal X})^2|,
\end{array}
  \label{Sympinv2}
\end{eqnarray}
where ~$\hat{  }$ stands for the codeterminant.
\begin{figure}[h]
 \begin{center}
\epsfxsize=0.5\textwidth \epsfbox{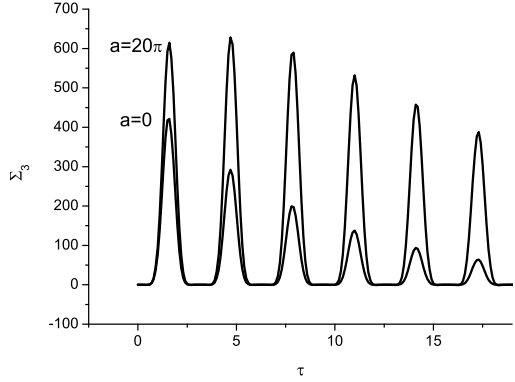}
\end{center}
\caption{The temporal evolution of the three-body uncertainty $\Sigma_3$ is shown.
 The initial condition is a two mode squeezed state with $r=0.2$ and a ground state of Chris.  Other parameters are $\gamma=0.005$,~$\Lambda=50$.
 $\Sigma_3\geq 0$ indicates that the uncertainty relation is always satisfied.
\label{fig5}}
\end{figure}
In Fig. 5, $\Sigma_3$ is plotted as a function of time.
$\Sigma_3\geq 0$ throughout the entire evolution indicates that the uncertainty relation is always satisfied.
The effect of acceleration is again similar to the effect of temperature which  increases the uncertainty.

Under the partial transpose on the first variable,
$|{A}_{12}|$,$|{A}_{13}|$,$|\hat{A}_{12}|$,$|\hat{A}_{13}|$ change signs
and the uncertainty after the partial transpose $\tilde{\Sigma}_3$ becomes
\begin{eqnarray}
\tilde{\Sigma}_3\equiv
-\frac{1}{64}+\frac{\tilde{\Delta}_{1}^{3}}{16}-\frac{\tilde{\Delta}_{2}^{3}}{4}
+\tilde{\Delta}_{3}^{3},
  \label{SigmaPT}
   \end{eqnarray}
   where
\begin{eqnarray}
\begin{array}{lll}
\tilde{\Delta}_{1}^{3} &=& |D_1|+|D_2|+|D_3|-2|A_{12}|-2|A_{13}
|+2|A_{23}|,\\ \nonumber
\tilde{\Delta}_{2}^{3} &=& |\hat{D}_1|+|\hat{D}_2|+|\hat{D}_3|-2|\hat{A}_{12}|-2|\hat{A}_{13}|+2|\hat{A}_{23}|,\\ \nonumber
\tilde{\Delta}_{3}^{3} &=& |(\Delta {\cal X})^2|.
\end{array}
  \label{Sympinv3}
\end{eqnarray}

From (37),
\begin{eqnarray}
\tilde{\Sigma}_3&=&
 (\lambda_{1}^2-\frac{1}{4})...(\lambda_{3}^2-\frac{1}{4}).
  \label{SigmaPTsimp}
   \end{eqnarray}
For separable states, each term on the right hand side is nonnegative,
   thus $\tilde{\Sigma}_3\geq 0$ follows.
For our interest of $(1+2)$ mode bipartite system, or more generally,
$(1+n)$ bipartite system, $\tilde{\Sigma}_{(1+n)}\geq 0$ is shown to give a necessary and sufficient condition for separability\cite{Serafini06}.

Three-mode entanglement for continuous variables can be categorized into the following sets\cite{GKLC01} based on the values of the bipartite entanglement measure $E_{i}$ with $i=A(BC),(AB)C,(AC)B$ ($E_{A(BC)}$ is bipartite entanglement between Alice and Bob-Chris together):
(1) fully entangled states (none of $E_{i}$ is vanishing.),
(2) one-mode biseparable states (only one of $E_{i}$ is vanishing.),
(3) two-mode biseparable states (only two of $E_{i}$ are vanishing.),
(4) three-mode biseparable states (all of $E_{i}$ are vanishing but these states
cannot be written as a mixture of tripartite product states),
(5) fully separable states (mixture of tripartite product states).

For bipartite entanglement for three qubits, the square of the concurrence $C$ is known to
satisfy the following monogamy inequality\cite{CKW00}
\begin{eqnarray}
C_{A(BC)}^2 \geq C_{AB}^2 + C_{AC}^2,
  \label{monogamy}
   \end{eqnarray}
   where $C_{AB}$ is the concurrence between $A$ and $B$, etc.
The inequality also holds for Gaussian states if we take
the square of the logarithmic negativity\cite{ASI06}.
Genuine tripartite entanglement for continuous variables $E_{ABC}$ can be defined by minimizing the difference between the left-hand and the right-hand side for all possible combinations of three subsystems, namely,
\begin{eqnarray}
E_{ABC} = \mbox{min}_{ijk} \left\{E_{i(jk)} - E_{ij} - E_{ik}\right\}.
  \label{3tangle}
   \end{eqnarray}
Note that the monogamy inequality guarantees $E_{ABC}\geq 0$.
   In Fig. 6, the temporal evolution of bipartite entanglement $E_{AB}$,$E_{AC}$,$E_{BC}$ are shown.
The initial state is chosen to be the product state of a two-mode
squeezed state of Alice and Bob and a coherent state of Chris.
Thus initially $E_{AC}=E_{BC}=0$. As time evolves, decoherence due
to the interaction with the environment\cite{Dec96} causes
$E_{AB}$ to damp out. The interaction with the environment induces
entanglement between Alice and Chris's state, which evolves
periodically in time and some of entanglement in $E_{AB}$ tranfers
to $E_{AC}$. There is no entanglement between Bob and Chris's
state during the entire period since there is neither initial nor
interaction-induced entanglement between them. Note a dual role of
the environment. It is both the source of multipartite
entanglement between observers and of decoherence that decreases
entanglement. The former causes initial increase and the latter
causes damping of entanglement in time.
 The similar behavior is already seen in two static Brownian oscillators\cite{Shiokawa08}.
In Fig. 7, the temporal evolution of $E_{ABC}$ is shown.
The initial state is the same as in Fig. 6, a product state of a two-mode squeezed state of Alice and Bob and a ground state of Chris.
Thus initially there is no genuine tripartite entanglement. The nonlocal vacuum generates entanglement among three variables. The same vacuum also induces decoherence
which eventually washes away all entanglement.
The Unruh effect suppresses both bipartite and tripartite entanglement.
\begin{figure}[h]
 \begin{center}
\epsfxsize=0.4\textwidth \epsfbox{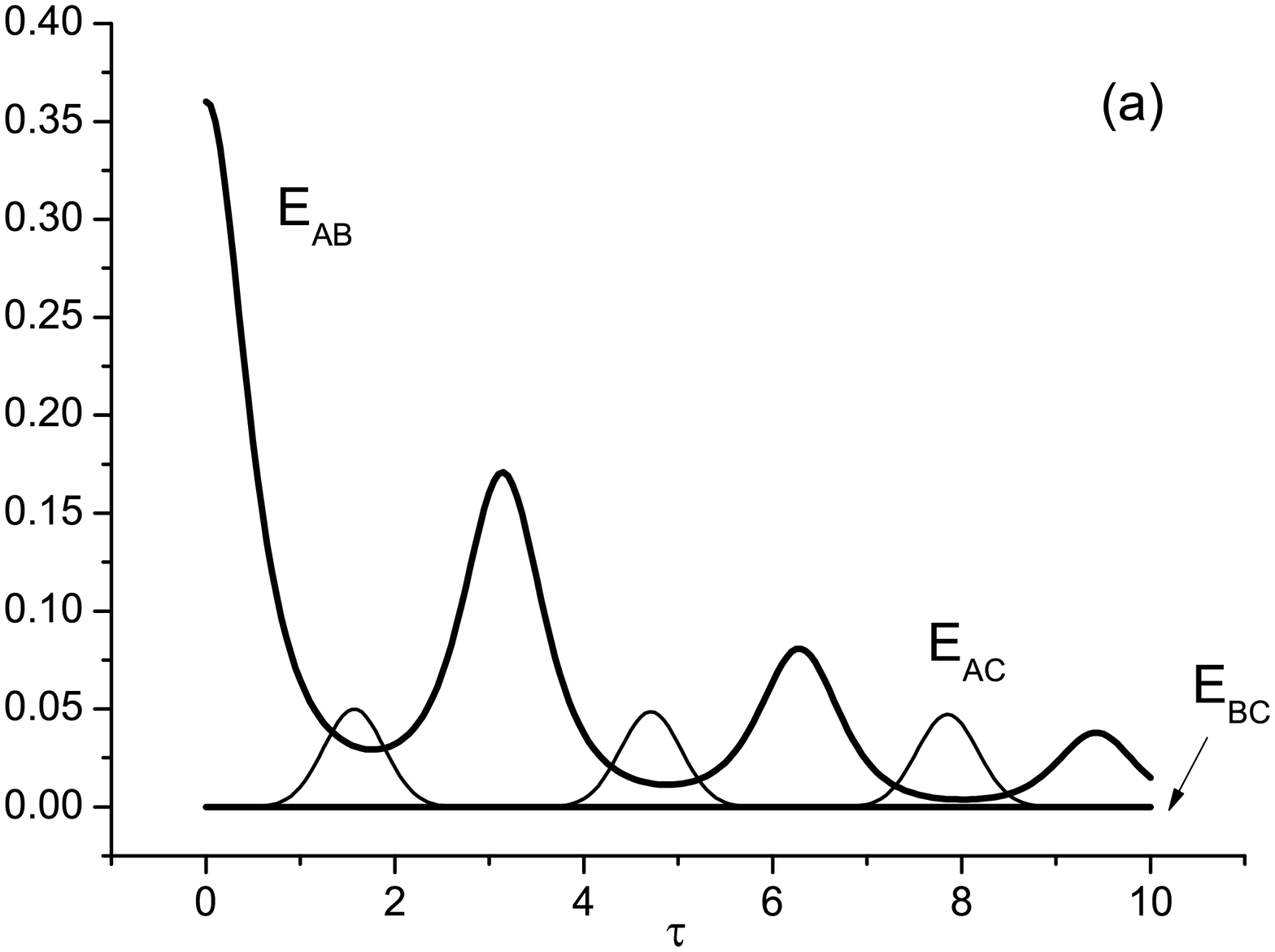}
\epsfxsize=0.4\textwidth \epsfbox{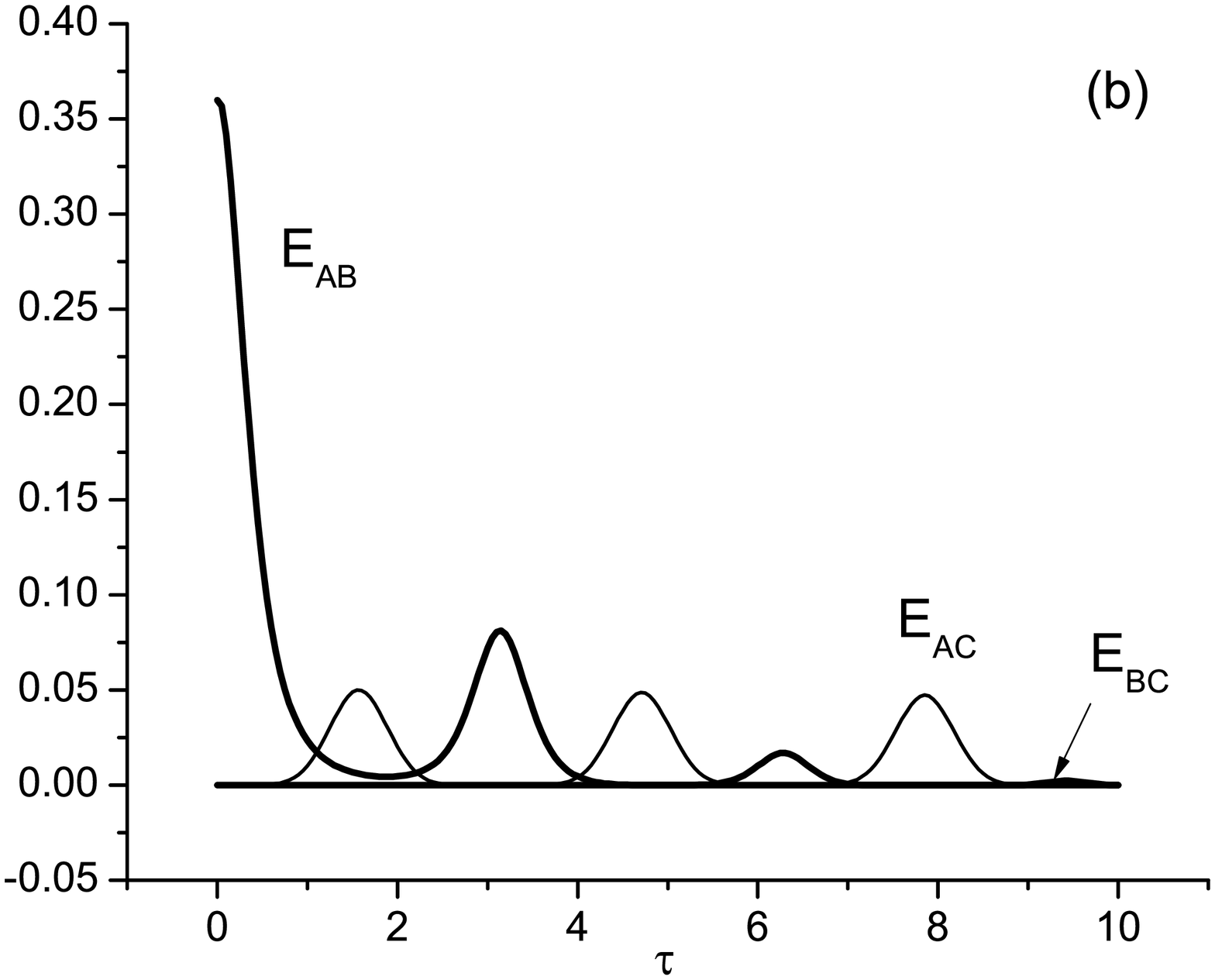}
\end{center}
\caption{The temporal evolution of bipartite entanglement $E_{AB}$, $E_{AC}$, and $E_{BC}$ are plotted as a function of time. The initial condition is a product state of a two-mode squeezed state of Alice and Bob with $r=0.3$ and a ground state of Chris. $a=0$ in Fig.6(a) and $a=50\pi$ in Fig.6(b). Other
 parameters are
$\gamma=0.001$,~$\Lambda=50$.
\label{fig6}}
\end{figure}
\begin{figure}[h]
 \begin{center}
\epsfxsize=0.4\textwidth \epsfbox{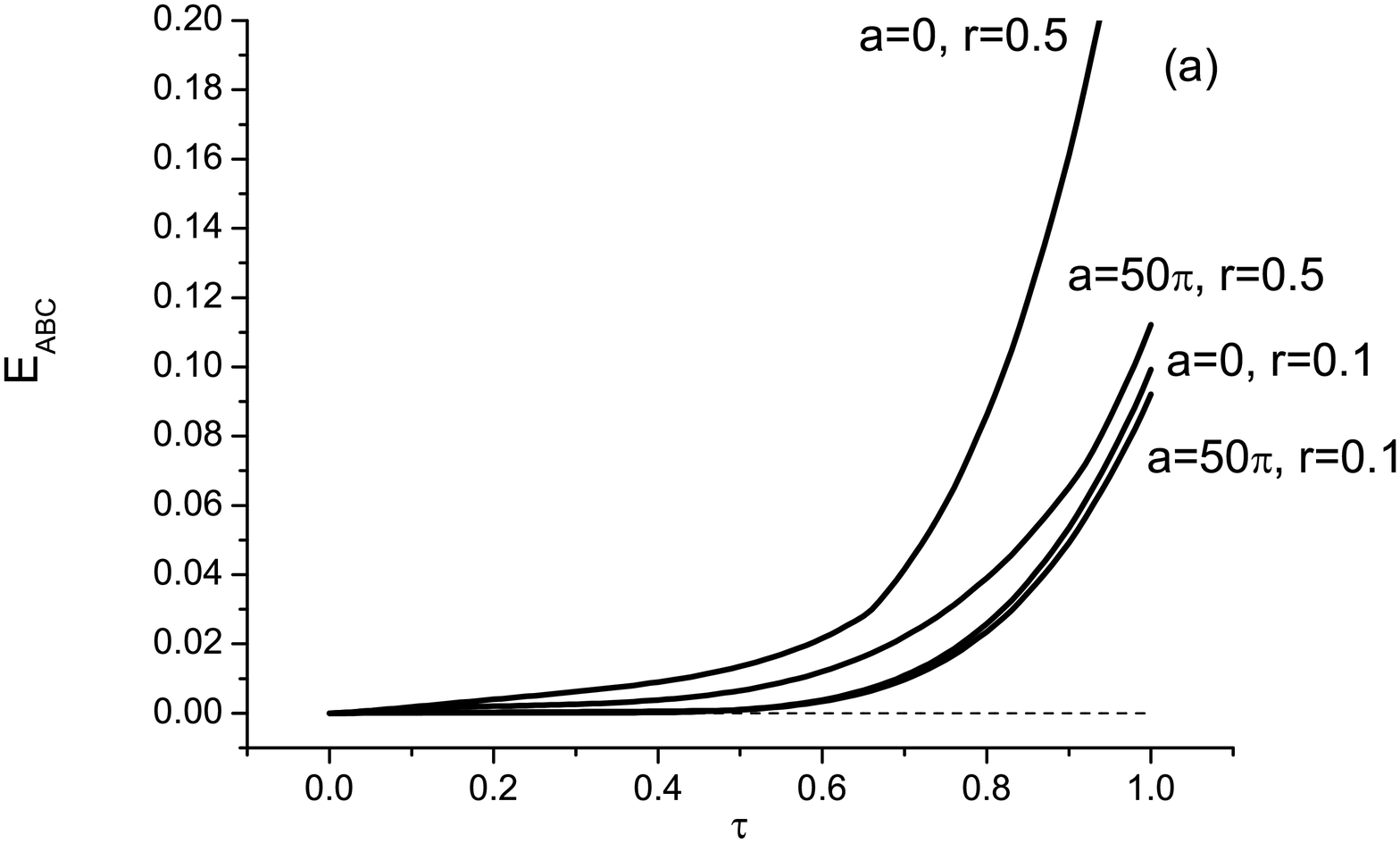}
\epsfxsize=0.4\textwidth \epsfbox{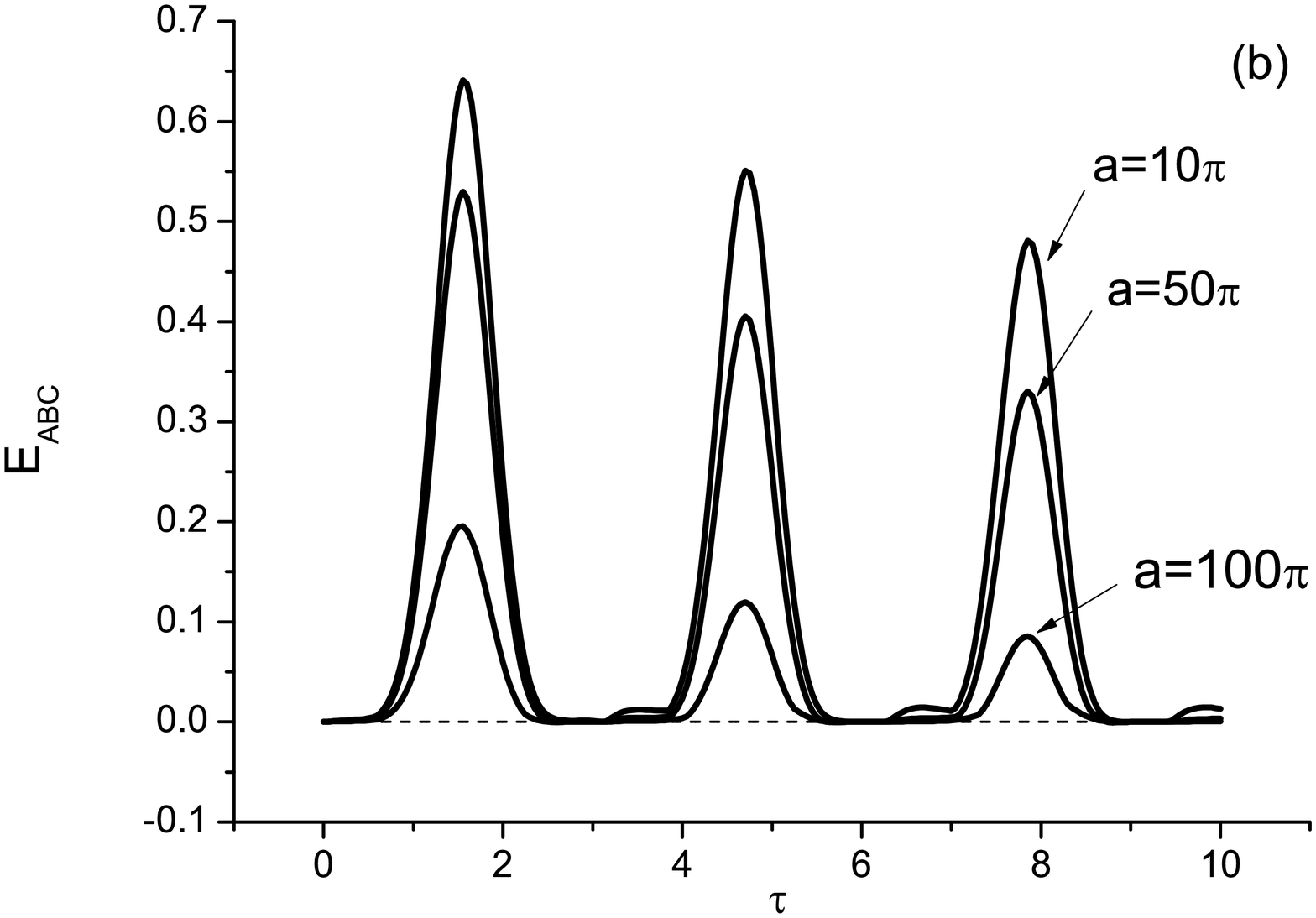}
\end{center}
\caption{The temporal evolution of the genuine tripartite entanglement $E_{ABC}$ is plotted as a function of time. The initial condition and unshown parameters are the same as in Fig. 6. The short time behavior is shown in (a). While squeezing of the initial state enhances $E_{ABC}$, acceleration suppresses it. The long time behavior in (b) is damped oscillatory.
\label{fig7}}
\end{figure}

\section{Quantum teleportation with an accelerated observer} \label{QTEL}

Many interacting quantum systems can be simulated with a proper
design of quantum circuits\cite{NielsenChuang00},  where the
interaction among subsystems are mimicked by quantum gates. Along
with the properly chosen measurement and post-selection scheme, we
can simulate many interesting systems under nontrivial situations.
In \cite{MaldaneceHorowitz04}, similarity between information
escaping from a black hole and a quantum teleportation protocol
was pointed out. In the presence of the generic interaction
between collapsing matter and a Hawking particle, however,
unitarity cannot be completely
rescued\cite{GottesmanPreskill04}. Encoding quantum states with quantum error
correction may still recover the full information.

In these works, only speculations about the possible quantum process in and around a black hole are given based on the analogies with quantum circuits. No analysis on the curved spacetime was provided.
It is important to note that qubits studied there are in a flat spacetime without event horizon while
information loss is predicted in quantum theory in curved spacetimes with event horizon that hides information.

Since our model can simulate the quantum dynamics of many observers moving in an arbitrary fashion, it provides a good setting to study the Unruh effect on various quantum protocols.
From the equivalence principle, the acceleration due to gravitational force can be
locally transformed away by the appropriate
coordinate transformation. In the absence of other forces, the observer then follows a geodesic in a curved spacetime.
The Unruh effect on a constantly accelerated observer can be identified as the effect on the static observer in Rindler spacetime.
Since near the Rindler horizon, the Rindler spacetime looks similar to the Schwartzchild black hole spacetime, this is equivalent to studying the Hawking effect on the quantum dynamics of a static observer near a black hole event horizon.
\begin{figure}
 \begin{center}
\epsfxsize=0.4\textwidth \epsfbox{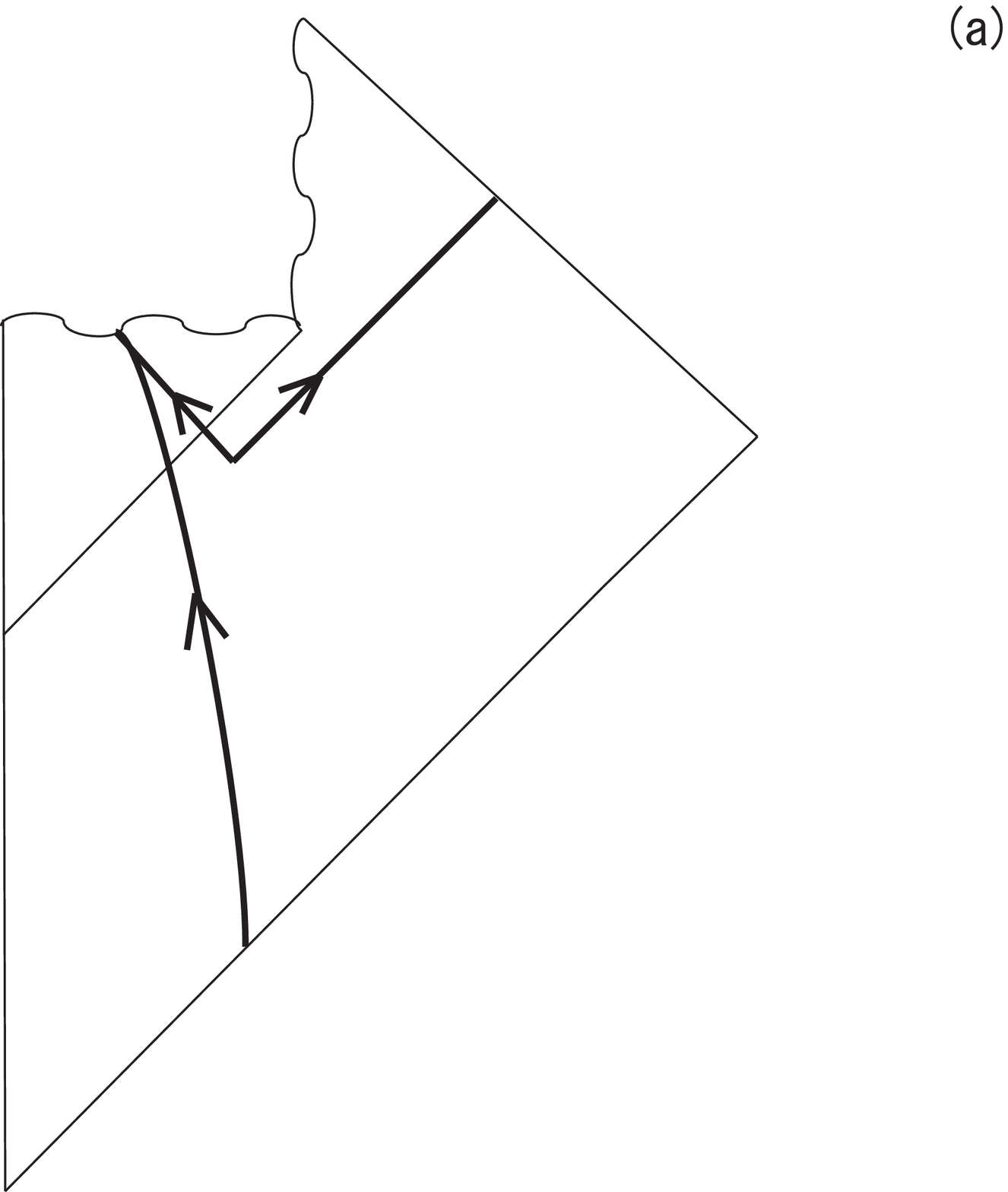}
\epsfxsize=0.33\textwidth \epsfbox{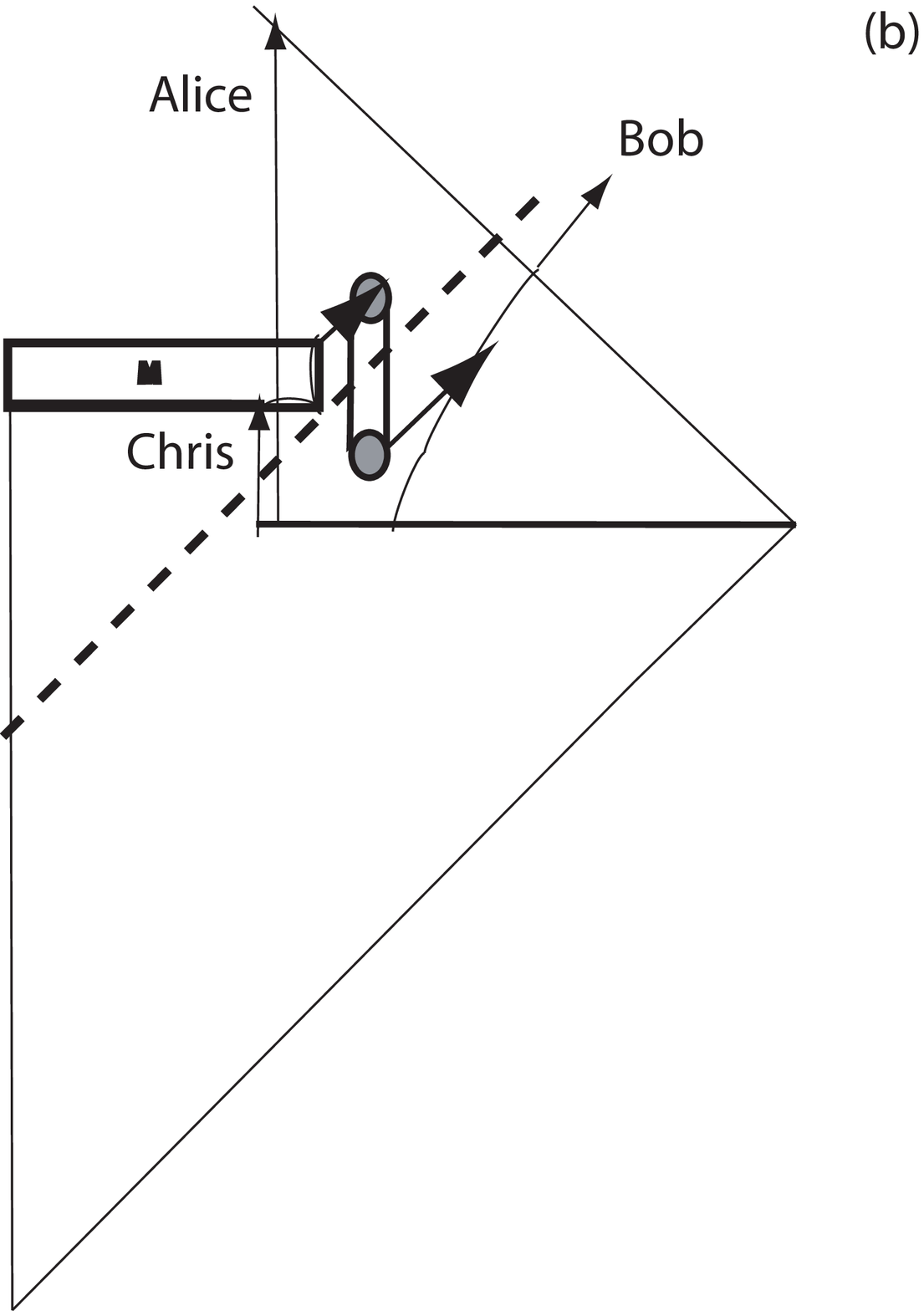}
\end{center}
\caption{Penrose diagrams of (a) Schwartzchild spacetime with a falling matter
and a Hawking pair and (b) Rindler spacetime with our quantum teleportation  scheme. Shaded region contains a pair of wormholes through which we can retrieve
measurement results beyond the Rindler horizon.  \label{fig8}}
\end{figure}
In Fig. 8, our quantum teleportation scheme in (b) is compared to
the scenario in \cite{MaldaneceHorowitz04} of information escaping
from a black hole in (a). In both figures, two parties are falling
inside of the horizon and hit one region while the other party is
moving toward future without falling inside of the horizon. The
role of singularity in (a) is viewed as an effectively boundary
condition, while it is replaced by the measurement performed at
some time in (b). If Bob is under constant acceleration
indefinitely, once Alice passes the event horizon, we need an
additional mechanism to send the measurement result to Bob for the
teleportation to be successful. In the figure, a pair of wormholes
are used to retrieve the measurement result. In
\cite{MaldaneceHorowitz04}, this was simply assumed as pointed out
in \cite{GottesmanPreskill04}. In the present work, we view our
scheme as a simulation of the process possibly occurring around a
black hole and are only interested in the process until the
measurement is performed. Bob can stop accelerating eventually and
receive the message from Alice. This is indeed the situation
analogous to a decaying black hole\cite{CarlitzWiley87}.
Although our system does not include gravitational interaction between
observers, as we saw in Sec. 3, the background field induces the effective interaction and entanglement between them.

Quantum teleportation with continuous variables based on \cite{BraunsteinKimble98} have been demonstrated
experimentally\cite{Furusawa:98}.
Initially Alice and Bob share a two-mode squeezed state. The mode
carried by Alice is mixed with Chris's state prepared in a coherent state through the $50$ \% beam-splitter. Their state is
measured by the homodyne detector and the result is sent to Bob
through the classical channel. Bob's state is shifted
according to the measurement result in order to reproduce the same
state as Chris's.

Now let us first discuss our problem, instead of measuring the state of Alice and Chris, by  tracing over all states for Alice and Chris at some time in the future\cite{FeyVer63}.
Since we are concerned with the Unruh effect on the quantum teleportation protocol,
first we consider the influence of the environment only on Alice and Bob's
state. Then Chris's state remains in a pure state and the density
matrix is the direct product of
Chris's and Alice and Bob's  state as $| \alpha \rangle_C \hat{\rho}_{AB}\langle
\alpha|_C$, where $\hat{\rho}_{AB}$ can be a mixed state density matrix.
The fidelity $F$ is an overlap
between Bob's final state and Chris's initial state given by
\begin{eqnarray}
F&=& _B\langle \alpha | {\mbox Tr}_{AC} \left[ |\alpha \rangle_C
\hat{\rho}_{AB}~_C\langle \alpha |\right] | \alpha \rangle_B \\
&=&\frac{2}{D^{(0)1/2}} \exp\left[
-\frac{\Gamma_1^{(0)}\alpha^2}{D^{(0)}}
-\frac{\bar{\Gamma}_1^{(0)}\bar{\alpha}^2}{D^{(0)}}
-\frac{2\Gamma_2^{(0)}}{D^{(0)}}|\alpha|^2 \right], \nonumber
\label{FNM}
\end{eqnarray}
where $\Gamma_1^{(0)}=(\Delta {\cal X})^2_{11}-(\Delta {\cal X})^2_{22}-2i\Delta {\cal X}^2_{12}$ and $\Gamma_2^{(0)}=(\Delta {\cal X})^2_{11}+(\Delta {\cal X})^2_{22}+1$
and $D^{(0)}=\Gamma_2^{(0)2}-|\Gamma_1^{(0)}|^2$.
 It is easy to see that the fidelity for this process gives an one-mode $Q$ distribution function for Bob's state.
The more Alice and Bob's quantum state is squeezed, the more it is entangled.
Then after tracing out Alice's degree of freedom, Bob's state
  is more mixed and the less fidelity results.
  This is opposite to what we expect in the teleportation scheme where large entanglement between Alice and Bob yields high fidelity in the teleported state as we will see below.

    The fidelity takes
its largest value at the center $\alpha=0$ as $F_{max} = 2D^{(0)-1/2}$.
After a long time, Bob's state thermalizes at the Unruh temperature.
 The off-diagonal components of the correlation matrix will vanish
and, in the weak coupling limit, their diagonal components are given by $(\Delta {\cal X})^2_{11}=(\Delta {\cal X})^2_{22}=\coth(\pi\Omega/a)/2$. This gives  $D^{(0)}=(\coth(\pi\Omega/a)+1)^2$ and
 the long time limit of $F$ can be written as a function of acceleration as
\begin{eqnarray}
F(\infty)&=& \frac{2}{\coth(\pi \Omega/a)+1} \exp\left[
-\frac{2}{\coth(\pi\Omega/a)+1}|\alpha|^2 \right].
\label{FNMlongt}
\end{eqnarray}
We see that this is independent of the initial squeezing.
The peak value $F_{max}(\infty)=2/\left[ \coth(\pi \Omega/a)+1 \right]$ takes from $1$ to $0$ as the acceleration
$a$ varies from $0$ to $\infty$. We see that $F_{max}$ measures the overlap between Bob's and the ground state.
For the zero acceleration case, $a=0$, $F_{max}(\infty)=1$, thus there is no loss of information.
\begin{figure}[hh]
 \begin{center}
\epsfxsize=0.4\textwidth \epsfbox{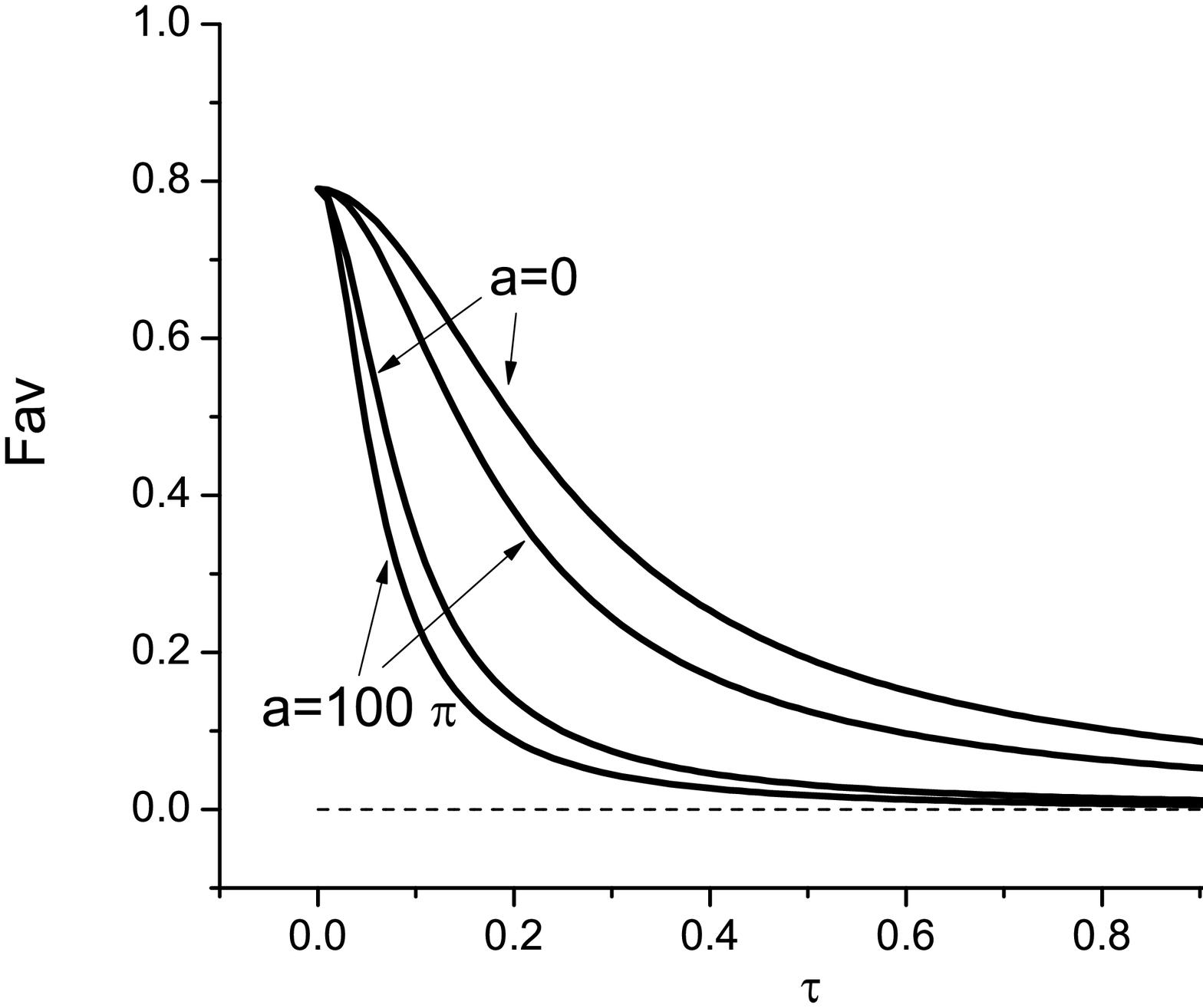}
\epsfxsize=0.4\textwidth \epsfbox{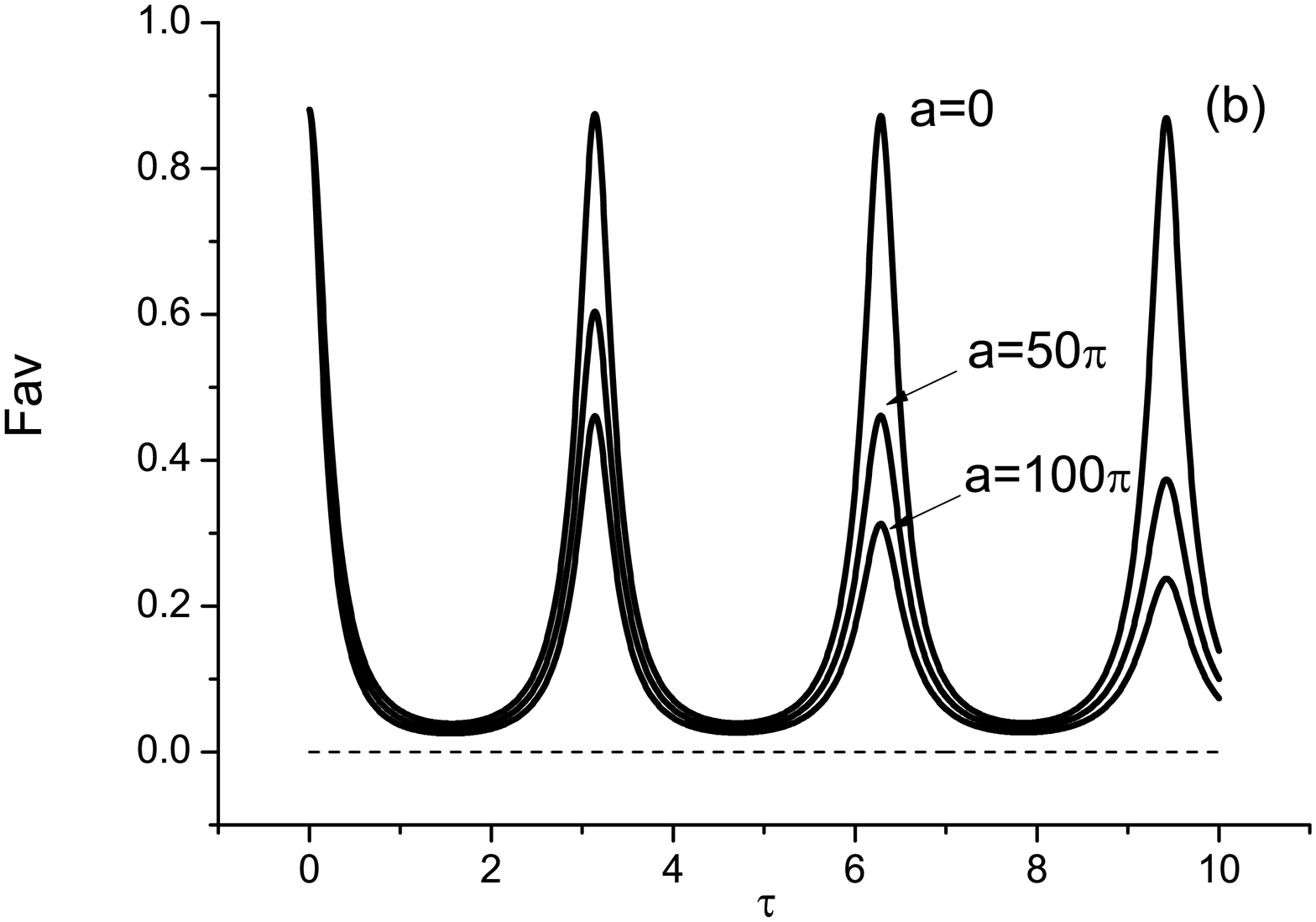}
\end{center}
\caption{The temporal evolution of the fidelity in (\ref{FNM}) is shown.
 The initial state of Alice and Bob is a two mode squeezed state with $r=0.2$.  Other parameters are $\Lambda=300$. $\gamma=0.002$ (right).
 The short time behavior is shown in (a). Increase in system-bath coupling and  acceleration both suppresses the fidelity. The long time behavior in (b) is damped oscillatory
 similarly to the temporal behavior of entanglement.
\label{fig9}}
\end{figure}
In Fig. 9, the time evolution of $F$ is plotted.
    The initial decay (a) is followed by the oscillations (b)
with the fundamental oscillator frequency $\Omega$. We see that the acceleration suppresses fidelity similarly to the effect of temperature.

In the scheme of quantum teleportation for qubits or continuous variables,
after Chris's state is mixed with Alice's,
their states are measured and the measurement result is send to Bob in a distance.
In the continuous variable quantum teleportation, the homodyne measurement
on Alice and Chris's state projects their state to the maximally entangled EPR state.
The imprecision involved in the measurement can be included, for instance, by making Gaussian smearing of the EPR state. Here
 we take into account the finite uncertainty in the projected state by considering a two-mode squeezed final state with a squeezing parameter $r_2$.
We denote the measurement result as a complex number $\beta=\beta_R+ i \beta_I$, where $\beta_R$ and $\beta_I$ are both real numbers.
Thus after the measurement, the original quantum state of Alice and Chris is projected to
\begin{eqnarray}
|\beta\rangle_{AC}&=&\cosh^{-1}r_2 \sum_{n=0}^{\infty} \tanh^{n}r_2
\hat{D}(\beta)|n \rangle_A|n \rangle_C, \label{stateafterHom}
\end{eqnarray}
where the translation operator $\hat{D}(\beta)$ is given by
\begin{eqnarray}
\hat{D}(\beta) &=& e^{\sqrt{2}i \beta_I \hat{x}_C - \sqrt{2}i \beta_R
\hat{p}_C}.
\end{eqnarray}
In the limit $r_2\rightarrow \infty$, $|\beta\rangle_{AC}$ becomes the maximally-entangled EPR state.

After the measurement of Alice and Chris's state,
the density matrix for Bob becomes
\begin{eqnarray}
 \hat{\rho}_{B} = N_{B} {}_{AC}\langle \beta | \alpha \rangle_C \hat{\rho}_{AB}~_C\langle \alpha |\beta \rangle_{AC},
\label{BobDM}
\end{eqnarray}
where $N_B=\cosh(r_1-r_2)\cosh(r_1+r_2)/\pi$ is a normalization constant.
With this normalization, $P(\beta)={\mbox Tr}_B \hat{\rho}_{B}$ is equal to the probability to obtain the measurement result $\beta$.
Bob's state is shifted depending on the measurement result $\beta$
so that the final state of Bob is $ \hat{\rho}_{out} = \hat{D}(\beta)
\hat{\rho}_{B}\hat{D}^{\dagger}(\beta)$.
 The teleportation fidelity $F$ is an overlap
between Bob's final state $ \hat{\rho}_{out}$ and Chris's initial state
$| \alpha_C \rangle$ and given by
\begin{eqnarray}
F=\frac{_B\langle \alpha | \hat{\rho}_{out} | \alpha \rangle_B}{P(\beta)}.
               \label{Fidel}
\end{eqnarray}

Suppose we do not make post-selection following the measurement, the fidelity becomes, instead of (\ref{Fidel}),
\begin{eqnarray}
F=\frac{_B\langle \alpha | \hat{\rho}_{B} | \alpha \rangle_B}{P(\beta)}.
               \label{FnoPS}
\end{eqnarray}
In this case, the averaged fidelity over all measurement outcome $\beta$ is
\begin{eqnarray}
F_{av}= \int d^2\beta P(\beta) F(\beta) = \int d^2\beta _B\langle
\alpha | \hat{\rho}_{B} | \alpha \rangle_B. \label{FavnoPS}
\end{eqnarray}
Similarly to (\ref{stateafterHom}) we write the initial two-mode squeezed state (\ref{WEPR})
in the Fock space representation as
\begin{eqnarray}
|{AB}\rangle&=&\cosh^{-1}r_1 \sum_{n=0}^{\infty} \tanh^{n}r_1
|n \rangle_A|n \rangle_B. \label{ABinitial}
\end{eqnarray}
In the absence of the environment, from (\ref{stateafterHom}),
we obtain $F_{av} = (\cosh r_1)^{-2} \exp\left[-|\alpha|^2/\cosh^{2}r_1\right]$
in the $r_2\rightarrow \infty$ limit.
Then $F_{av}\rightarrow 0$ as initial squeezing $r_1$ gets larger.
This tells us that quantum teleportation cannot be successful without post-selection.

On the other hand, with post selection, the average fidelity can be shown to be
\begin{eqnarray}
F_{av}= \frac{1+\tanh r_1}{2}. \label{FavPS}
\end{eqnarray}
For a coherent state $r_1$, $F_{av}=1/2$ and
for any positive squeezing, $F_{av}>1/2$.
Since $F_{av}=1/2$ is a maximal value achieved by the classical method\cite{BraunsteinFuchsKimble00}.
We see that the initial entanglement in Alice and Bob's state
is essential for quantum teleportation.

Now we consider the effect of the environment on the performance of teleportation.
In our situation, the environment acts continuously on the system
so that the final density matrix of Bob depends on how we send the measurement
result of Alice and Chris to Bob.
In order to eliminate this ambiguity, here we consider the effect of the environment
up to the point when Alice and Chris's state is measured.
We introduce the complex variables $\Delta_i~(i=1,...,6)$
to simplify the expression as follows:
\begin{eqnarray}
\Delta_1&=&(\Delta {\cal X})^2_{11}-(\Delta {\cal X})^2_{22}+2i(\Delta {\cal X})^2_{12}, \nonumber\\
\Delta_2&=&(\Delta {\cal X})^2_{11}+(\Delta {\cal X})^2_{22}, \nonumber\\
\Delta_3&=&(\Delta {\cal X})^2_{33}-(\Delta {\cal X})^2_{44}+2i (\Delta {\cal X})^2_{34}, \nonumber\\
\Delta_4&=&(\Delta {\cal X})^2_{33}+(\Delta {\cal X})^2_{44}, \nonumber\\
\Delta_5&=&(\Delta {\cal X})^2_{13}-(\Delta {\cal X})^2_{24}+2i((\Delta {\cal X})^2_{14}+(\Delta {\cal X})^2_{23}), \nonumber\\
\Delta_6&=&(\Delta {\cal X})^2_{13}+(\Delta {\cal X})^2_{24}+2i((\Delta {\cal X})^2_{14}-(\Delta {\cal X})^2_{23}).
\end{eqnarray}
With a measurement outcome $\beta$, $_{B}\langle \alpha | \hat{\rho}_{out} | \alpha \rangle_B$
and $P(\beta)$ can be calculated straightforwardly.
Both can be written in the Gaussian form:
\begin{eqnarray}
_{B}\langle \alpha | \hat{\rho}_{out} | \alpha \rangle_B &=& N_1 \exp\left[-\Gamma_1(\alpha-\beta)^2  -\bar{\Gamma}_1(\bar{\alpha}-\bar{\beta})^2
-2\Gamma_2 |\alpha-\beta|^2 \right]  \nonumber
\end{eqnarray}
and
\begin{eqnarray}
P(\beta)&=& N_{P1} \exp\left[-\Gamma_3(\alpha-\beta)^2 -\bar{\Gamma}_3(\bar{\alpha}-\bar{\beta})^2
  -2\Gamma_4 |\alpha-\beta|^2 \right]. \label{PbetaHom}
\end{eqnarray}
The normalization factors and coefficients in exponents are both dependent on the correlation matrix and initial condition. As a result, they have the complicated time dependence as follows.
We will write $c_i=\cosh(2r_i)$ and $s_i=\sinh(2r_i)$ for $i=1,2$ below:
\begin{eqnarray}
N_1 &=& \frac{8N_B}{(d_2d_3)^{1/2}}, \nonumber\\
 \Gamma_1 &=&\frac{\bar{p} f_1^2 + p \bar{f}_2^2 + 2 q f_1 \bar{f}_2}{d_3 d_2^2
d_1^2}-\frac{s_2^2 \Delta_1}{d_2},\nonumber\\
 \Gamma_2 &=&\frac{\bar{p} f_1 f_2 + p
\bar{f}_1 \bar{f}_2 + q (|f_1|^2+|f_2|^2)}{d_3 d_2^2 d_1^2}-\frac{s_2^2
c_4}{d_2},\nonumber\\
  c_3 &=& c_2+\Delta_2,\nonumber\\
c_4 &=& c_2+\Delta_2-\frac{d_1}{c_2-1},\nonumber\\
b_1 &=& \bar{\Delta}_1 \Delta_5^2 +\Delta_1 \Delta_6^2 +2c_3\Delta_5 \Delta_6,\nonumber\\
b_2 &=& \bar{\Delta}_1 \Delta_5 \bar{\Delta}_6 + \Delta_1 \bar{\Delta}_5 \Delta_6 +c_3
(|\Delta_5|^2+|\Delta_6|^2),\nonumber\\
d_1 &=& (c_2+\Delta_2)^2-|\Delta_1|^2,\nonumber\\
 d_2 &=& s_2^4 + d_1 (c_2+1)^2 -2 s_2^2 (c_2+1)(c_2+\Delta_2),\nonumber\\
 d_3 &=& q^2-|p|^2, \nonumber\\
 d_4 &=&(d_1(\Delta_4+1)-b_2)^2-(d_1\Delta_3-b_1)^2, \nonumber\\
p &=& \Delta_3 -d_1^{-1}[ \bar{\Delta}_1 \Delta_5^2+\Delta_1 \Delta_6^2+2c_3\Delta_5 \Delta_6], \nonumber\\
&-&d_2^{-1}d_1^{-2}s_2^4[ (|\Delta_1|^2 + c_3(c_3+c_4))(\Delta_5^2 \bar{\Delta}_1+ \Delta_6^2 \Delta_1)
 + 2 \Delta_5 \Delta_6 ( |\Delta_1|^2 (2c_3+c_4)+c_3^2 c_4) ], \nonumber\\
q&=&\Delta_4+1-d_1^{-1}[\bar{\Delta}_1\Delta_5\bar{\Delta}_6+
\Delta_1\bar{\Delta}_5\Delta_6+c_3(|\Delta_5|^2+|\Delta_6|^2)]+  \nonumber\\
d_2^{-1}d_1^{-2}s_2^4&[& (|\Delta_1|^2 + c_3(c_3+c_4))(\Delta_5 \bar{\Delta}_1 \bar{\Delta}_6 + \bar{\Delta}_5\Delta_1 \Delta_6)
 +(|\Delta_1|^2(2c_3+c_4) +c_3^2 c_4)(|\Delta_5|^2+|\Delta_6|^2)], \nonumber\\
f_1 &=&  d_2 d_1 -s_2^3[\Delta_5(|\Delta_1|^2+c_3c_4)+\Delta_1\Delta_6 (c_3+c_4)],\nonumber\\
f_2 &=& -s_2^3 [ \Delta_5\bar{\Delta}_1(c_3+c_4)+\Delta_6  (|\Delta_1|^2+c_3c_4)],\nonumber\\
\Gamma_3 &=&d_1(d_1\Delta_3-b_1)/d_4,\nonumber\\
\Gamma_4 &=&d_1(d_1(\Delta_4+1)-b_2)/d_4, \nonumber\\
N_{P1} &=& \frac{4 c_3 N_B}{d_4^{1/2}}.
\label{VariousCoeff}
\end{eqnarray}

The averaged fidelity over all measurement outcome $\beta$ is
\begin{eqnarray}
F_{av}&=& \int d^2\beta P(\beta) F(\beta) = \int d^2\beta _B\langle
\alpha | \hat{\rho}_{out} | \alpha \rangle_B \nonumber\\ &=& \frac{4\pi N_B}{(d_2 d_3
d_5)^{1/2}},  \label{Fav}
\end{eqnarray}
where
\begin{eqnarray}
d_5 &=& \Gamma_2^2 - |\Gamma_1|^2 \label{G}.
\end{eqnarray}
\begin{figure}[h]
 \begin{center}
 \epsfxsize=0.4\textwidth \epsfbox{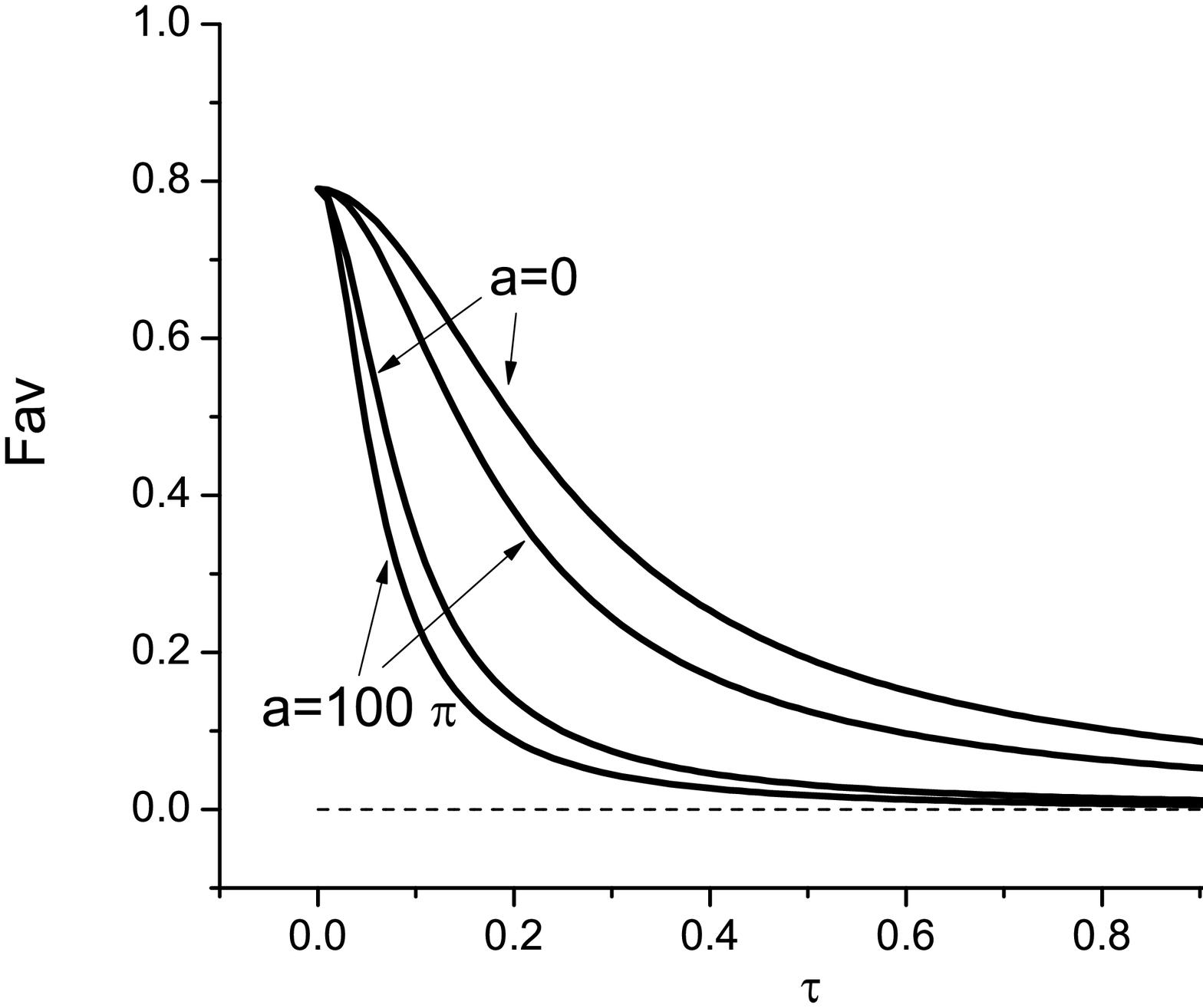}
\epsfxsize=0.4\textwidth \epsfbox{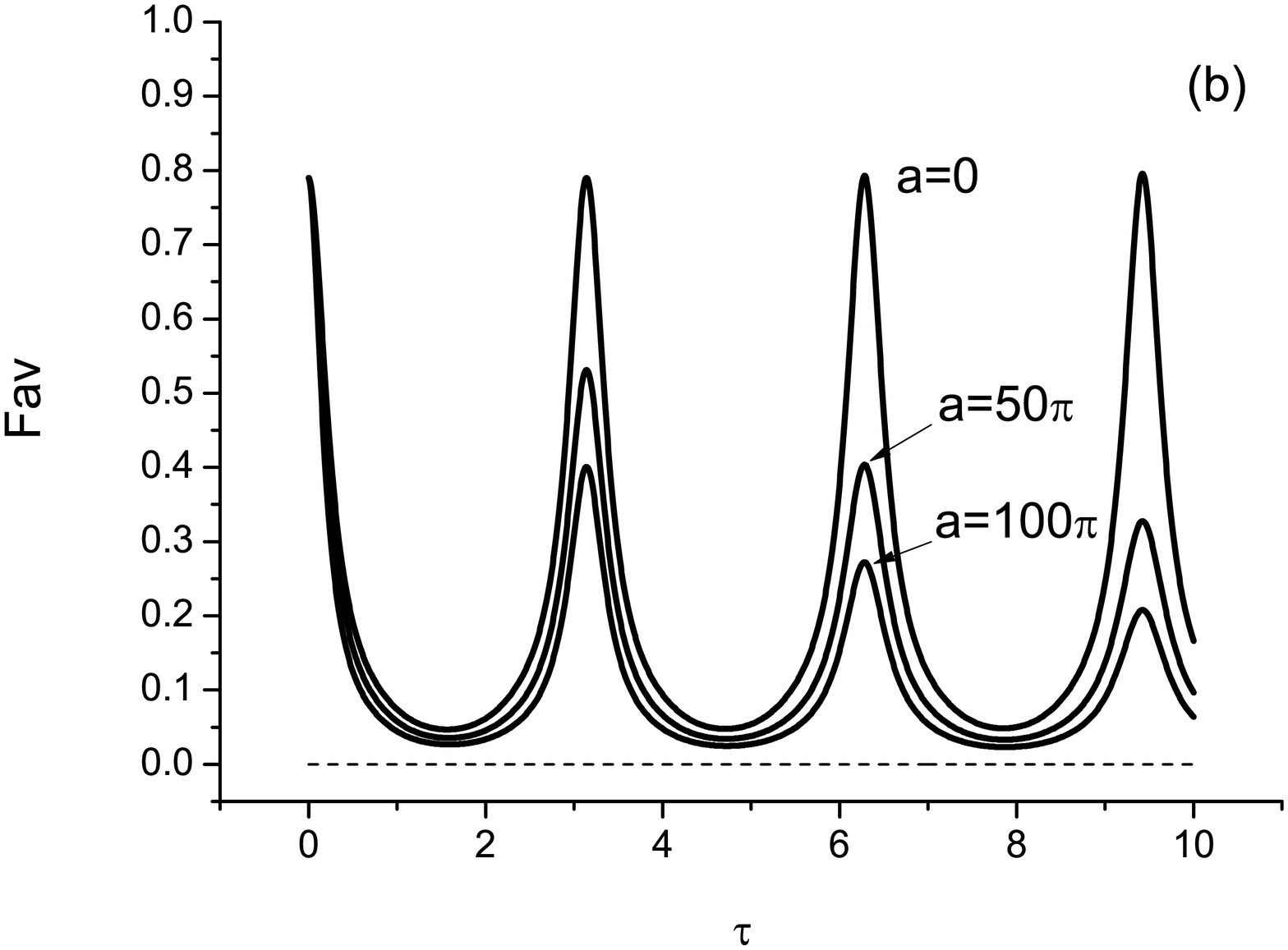}
\end{center}
\caption{The temporal evolution of the fidelity for the two mode squeezed final state with the squeezing parameter $r_2=1.0$ is shown.
 The initial state is a two mode squeezed state with the squeezing parameter $r_1=1.0$. $\gamma=0.002$, and $\Lambda=300$. The short time behavior in (a) shows that strong system-bath coupling and acceleration both suppress the fidelity. The long time behavior in (b) is damped oscillatory.
\label{fig10}}
\end{figure}
In Fig. 10, the time evolution of the averaged fidelity is plotted.
For the final state to be in a squeezed state with a finite squeezing parameter
$r_2$, the initial fidelity is not one. The short time decay rate is increased with
stronger system-field coupling and with larger acceleration of Bob as seen in Fig. 10(a).
The long time evolution exhibits damped oscillations.
The larger the acceleration, the larger the damping rate results.

The expressions in (\ref{VariousCoeff}) become much simpler if there is no interaction
with the environment. In this case, $\Delta_1,\Delta_3,\Delta_6 \rightarrow 0$ and
$\Delta_2,\Delta_4 \rightarrow c_1, \Delta_5\rightarrow -s_1$.
Other terms are
\begin{eqnarray}
p,f_2,\Gamma_1,\Gamma_3,b_1 \rightarrow 0
\end{eqnarray}
and
\begin{eqnarray}
c_3 &\rightarrow& c_1 + c_2, \nonumber\\
c_4 &\rightarrow& -\frac{(c_1+c_2)(c_1+1)}{c_2-1}, \nonumber\\
 d_2 &\rightarrow& (c_1+1)^2(c_2+1)^2,\nonumber\\
 q &\rightarrow& 2,\nonumber\\
d_3 &\rightarrow& 4,\nonumber\\
N_1 &\rightarrow& \frac{4 N_B}{(c_1+1)(c_2+1)},\nonumber\\
d_1 &\rightarrow& (c_1+c_2)^2,\nonumber\\
f_1 &\rightarrow& (c_1+c_2)^2 (c_1+1) (c_2+1) \left[(c_1+1)(c_2+1)+s_1 s_2\right],\nonumber\\
 \Gamma_2 &\rightarrow& 1 - \frac{s_1 s_2}{(c_1+1)(c_2+1)}, \nonumber\\
\Gamma_4
 &\rightarrow& \frac{c_1+c_2}{(c_1+1)(c_2+1)}, \nonumber\\
N_{P1}
&\rightarrow& \frac{4 N_B (c_1+c_2)}{(c_1+1)(c_2+1)}, \nonumber\\
d_4
&\rightarrow& (c_1+c_2)^2 (c_1+1)^2(c_2+1)^2,  \nonumber\\
b_2 &\rightarrow& (c_1+c_2) s_1^2.
\end{eqnarray}
From (\ref{PbetaHom}), we have
\begin{eqnarray}
_B\langle \alpha | \hat{\rho}_{out} | \alpha \rangle_B &=&
\frac{4N_B}{(c_1+1)(c_2+1)} e^{-2 |\alpha-\beta|^2 \Gamma_2} \label{Hom}
\end{eqnarray}
and
\begin{eqnarray}
P(\beta) &=& \frac{4N_B}{(c_1+1)(c_2+1)}e^{-2 |\alpha-\beta|^2 \Gamma_4}. \label{Pb0}
\end{eqnarray}
Then from (\ref{Fidel}),
\begin{eqnarray}
F&=& \exp\left[-2 |\alpha-\beta|^2 \frac{c_1 c_2 - s_1
s_2+1}{(c_1+1)(c_2+1)} \right] \label{Fnoenv}
\end{eqnarray}
and
\begin{eqnarray}
F_{av}&=& \frac{2 \cosh(r_1-r_2)\cosh(r_1+r_2)}{(c_1+1)(c_2+1) - s_1
s_2} \label{Favnoenv}
\end{eqnarray}
follow.
When the final projected state becomes maximally entangled, $r_2 \rightarrow \infty$, then $c_2,s_2 \rightarrow \infty$ and
\begin{eqnarray}
F_{av} &=& \frac{1+\tanh(r_1)}{2}. \label{Favnoenvr2infty}
\end{eqnarray}
Thus we reproduced Eq. (\ref{FavPS}).
 Furthermore, when the initial state of Alice and Bob becomes maximally entangled,  $r_1 \rightarrow \infty$, then $F_{av} \rightarrow 1$.
Thus there is no loss of fidelity in this limit.

In the proposed quantum teleportation scheme,
the state for Alice and Chris after measurement is assumed to be maximally entangled.
It is the Bell state for the qubit teleportation\cite{BBCJPW93} and the EPR state for the continuous
variable teleportation\cite{Vaidman94}.
The latter can be obtained by taking $r_2 \rightarrow \infty$ in (\ref{stateafterHom}) as
\begin{eqnarray}
|\beta\rangle_{AC} \sim \sum_{n=0}^{\infty}
\hat{D}(\beta)|n \rangle_A|n \rangle_C. \label{ACEPR}
\end{eqnarray}
This is the EPR state shifted by the displacement operator $\hat{D}(\beta)$.
In this limit, Bob's state after measurement becomes
\begin{eqnarray}
 \hat{\rho}_{B} = \frac{1}{\pi} \sum_{n,m=0}^{\infty}~_A\langle n |_C\langle n | \hat{D}^{\dagger}(\beta)|\alpha \rangle_C \hat{\rho}_{AB}~_C\langle \alpha |\hat{D}(\beta)|m \rangle_A|m \rangle_C.
\label{BobDMr2inf}
\end{eqnarray}
The elements in (\ref{PbetaHom}) become somewhat simpler as
\begin{eqnarray}
_B\langle \alpha | \hat{\rho}_{out} | \alpha \rangle_B &=& N_{2} \exp\left[-\Gamma_5(\alpha-\beta)^2  -\bar{\Gamma}_5(\bar{\alpha}-\bar{\beta})^2
-2\Gamma_6 |\alpha-\beta|^2 -\Gamma_7(\alpha-\beta)-\bar{\Gamma}_7(\bar{\alpha}-\bar{\beta})\right] \label{QHom} \nonumber
\end{eqnarray}
and
\begin{eqnarray}
P(\beta)&=& N_{P2} \exp\left[-\Gamma_8(\alpha-\beta)^2 -\bar{\Gamma}_8(\bar{\alpha}-\bar{\beta})^2
  -2\Gamma_9 |\alpha-\beta|^2 \right]. \nonumber\label{PbetaHom2}
\end{eqnarray}
The coefficients are
\begin{eqnarray}
N_{2} &=& \frac{4 e^{-\Gamma_{10}} D_1^{1/2}}{\pi D_2^{1/2}}, \nonumber\\
 d_5 &=& (\Delta_4+1)^2-|\Delta_3|^2,\nonumber\\
 d_6 &=& r^2 - |s|^2, \nonumber\\
 r &=&(\Delta_2+1)d_5-(\Delta_4+1)(|\Delta_5|^2+|\Delta_6|^2),   \nonumber\\
 s &=& \Delta_1 d_5-2(\Delta_4+1)\Delta_5 \bar{\Delta}_6,\nonumber\\
 u &=& d_5-(\Delta_4+1)\Delta_5, \nonumber\\
 v &=& -(\Delta_4+1)\bar{\Delta}_6, \nonumber\\
 w &=& -\bar{\Delta}_3\Delta_5 -\Delta_3 \bar{\Delta}_6,\nonumber\\
N_{P2}  &=& \frac{2}{\pi d_5^{1/2}} ,\nonumber\\\Gamma_5 &=&\frac{1}{d_5 d_6}
 \left[ \bar{s}u^2+s\bar{v}^2+2ru\bar{v} \right], \nonumber\\
 \Gamma_6 &=&\frac{1}{d_5 d_6}
 \left[ \bar{s}uv+s\bar{u}\bar{v}+r(|u|^2+|v|^2) \right]+\frac{\Delta_4+1}{d_5}, \nonumber\\
 \Gamma_7 &=&\frac{1}{d_5 d_6}
 \left[ \bar{s}uw+s\bar{v}\bar{w}+ru\bar{w}+r\bar{v}w\right]
 +\frac{\Delta_3}{d_5}, \nonumber\\
\Gamma_8 &=& \Delta_3 d_5^{-1},\nonumber\\
\Gamma_9 &=&  (\Delta_4+1) d_5^{-1}, \nonumber\\
\Gamma_{10} &=& \frac{1}{4 d_5 d_6}\left[ \bar{s}w^2+s\bar{w}^2+2r|w|^2 \right].
\label{VariousCoeffinf}
\end{eqnarray}
The averaged fidelity over all measurement outcome $\beta$ in this case is
\begin{eqnarray}
F_{av}&=& \int d^2\beta P(\beta) F(\beta) = \int d^2\beta _B\langle
\alpha | \hat{\rho}_{out} | \alpha \rangle_B \nonumber\\ &=& \frac{\pi N_{2}  e^{-\Gamma_{10}+\Gamma_{11}}}{2 d_7^{1/2}},  \label{Favmaxfinal}
\end{eqnarray}
where
\begin{eqnarray}
d_7 &=& \Gamma_6^2 - |\Gamma_5|^2 \nonumber\\
 \Gamma_{11} &=&\frac{1}{4D_3}\left( \bar{\Gamma}_5 \Gamma_7^2+\Gamma_5\bar{\Gamma}_7^2-2\Gamma_6|\Gamma_7|^2   \right).
\label{VariousCoeffinf2}
\end{eqnarray}
\begin{figure}[h]
 \begin{center}
 \epsfxsize=0.4\textwidth \epsfbox{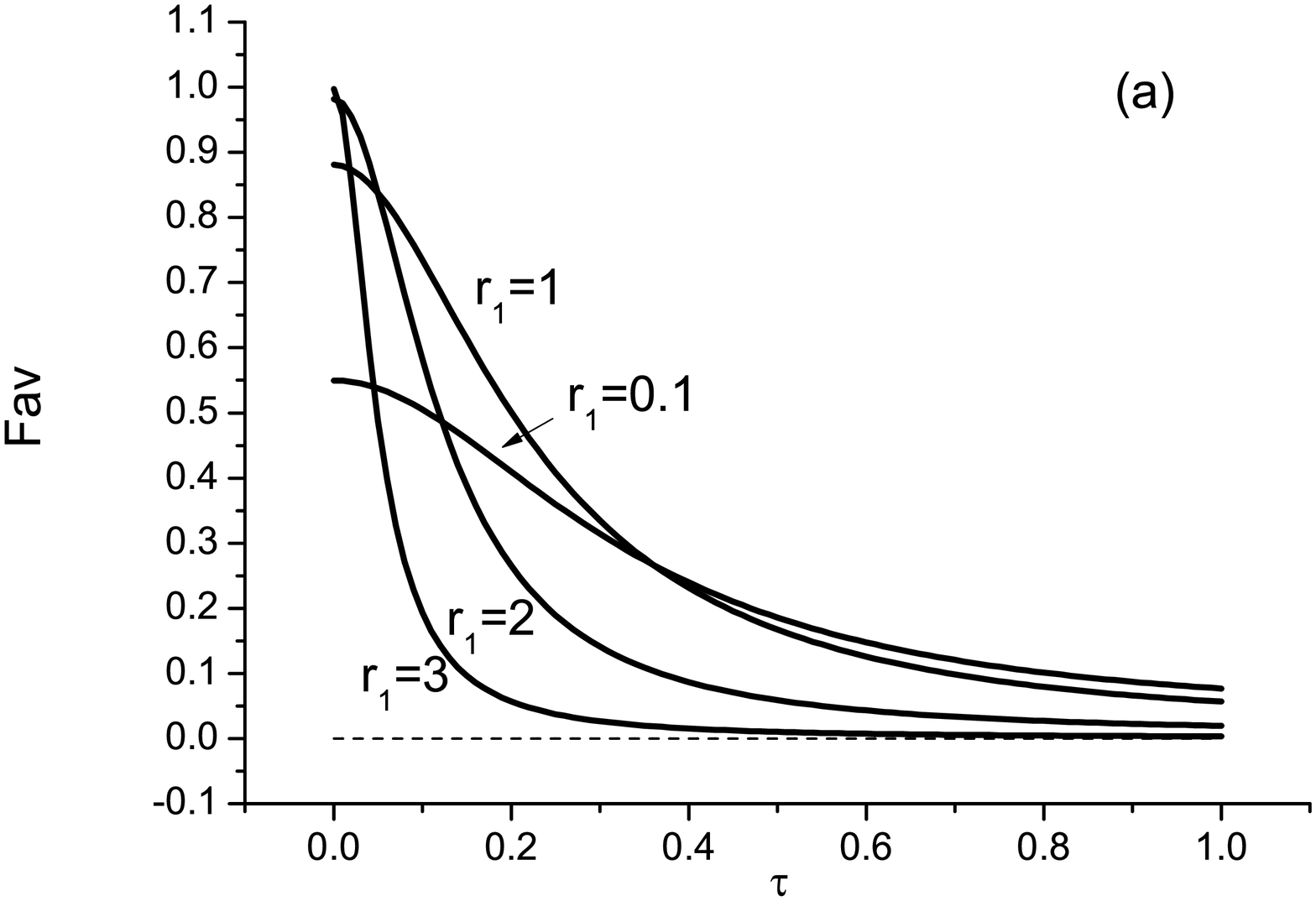}
\epsfxsize=0.4\textwidth \epsfbox{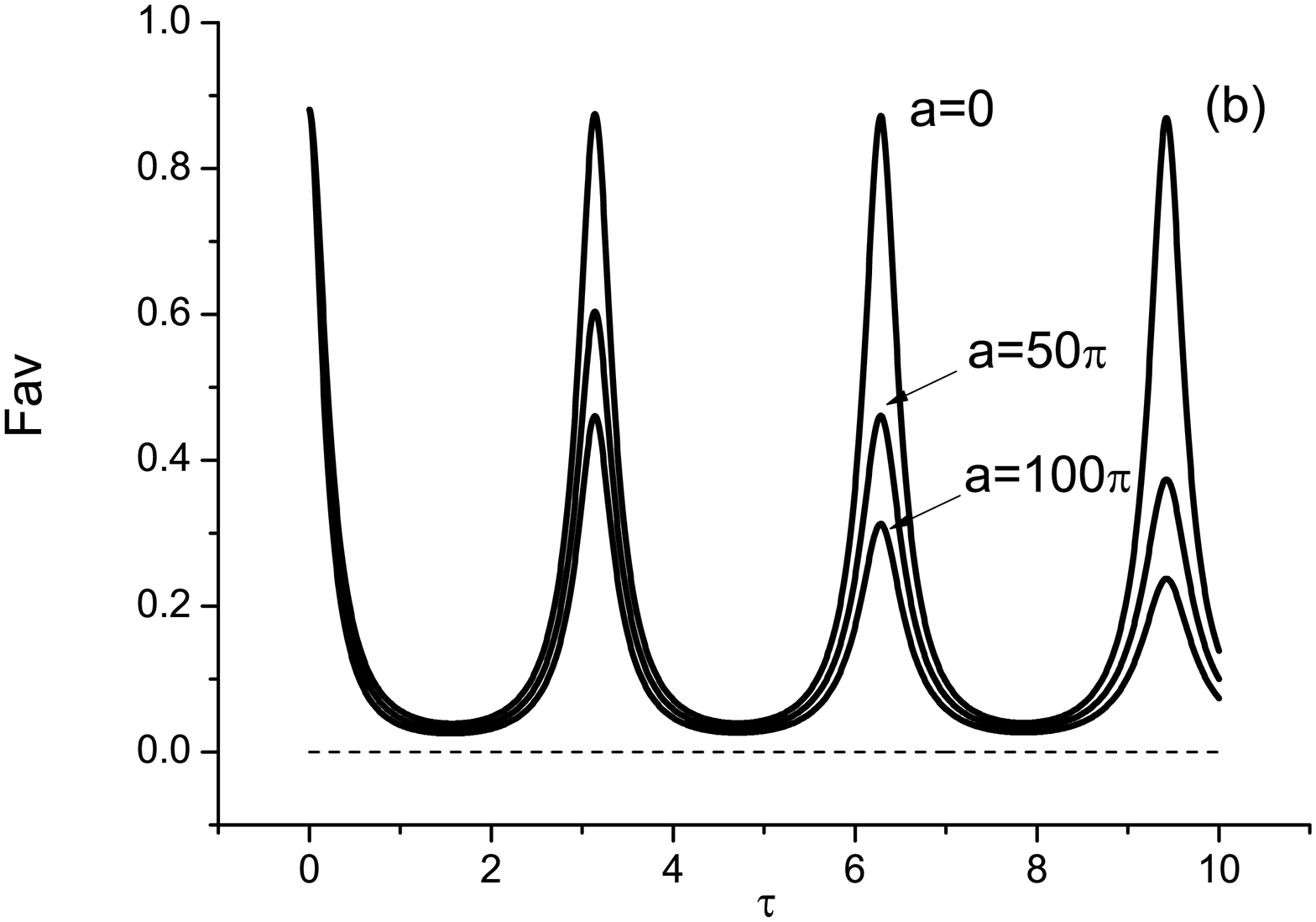}
\end{center}
\caption{The temporal evolution of the average fidelity for the maximally entangled final state is shown.
 The initial condition is a two mode squeezed state with the squeezing parameter $r_1$. The short time regime is shown in (a) with $a=0$. The long time regime
  is shown in (b) with $r_1=1.0$.  Other
 parameters are
$\gamma=0.002$,~$\Lambda=300$.
\label{fig11}}
\end{figure}
In Fig. 11, the time evolution of the average fidelity for the maximally entangled final state is plotted.
For this final state, the initial average fidelity reaches the unit value as
initial squeezing becomes infinitely large as seen in Fig. 11(a).
The long time evolution is again damped oscillatory with the damping rate enhanced
by acceleration as shown in Fig. 11(b).

In the absence of the environment,
\begin{eqnarray}
a,q,r,\Gamma_5,\Gamma_7,\Gamma_8,\Gamma_{10},\Gamma_{11} \rightarrow 0.
\end{eqnarray}
Other coefficients will be simplified accordingly as
\begin{eqnarray}
N_{2} &\rightarrow& \frac{2}{\pi(c_1+1)}, \nonumber\\
 \Gamma_6 &\rightarrow& 1-\frac{s_1}{c_1+1}, \nonumber\\
 d_5 &\rightarrow& (c_1+1)^2, \nonumber\\
 d_6 &\rightarrow& 4(c_1+1)^4, \nonumber\\
 d_7 &\rightarrow& \left(1-\frac{s_1}{c_1+1}\right)^2, \nonumber\\
 r&\rightarrow&2(c_1+1)^2,  \nonumber\\
 u &\rightarrow& 2(c_1+1)(c_1+s_1+1), \nonumber\\
 N_{P2}  &\rightarrow& \frac{2}{\pi(c_1+1)},\nonumber\\
\Gamma_9 &\rightarrow& \frac{1}{c_1+1}.
\label{VariousCoeffinfmaximalFS}
\end{eqnarray}
Thus we have
\begin{eqnarray}
_B\langle \alpha | \hat{\rho}_{out} | \alpha \rangle_B =
\frac{2}{\pi(c_1+1)} e^{-2 |\alpha-\beta|^2 \left(1-\frac{s_1}{c_1+1}\right) } \label{Hom0}
\end{eqnarray}
and
\begin{eqnarray}
P(\beta)= \frac{2}{\pi(c_1+1)}e^{-2 |\alpha-\beta|^2 \frac{1}{c_1+1}}. \label{Pb02}
\end{eqnarray}
The fidelity becomes
\begin{eqnarray}
F=\exp\left[-2 |\alpha-\beta|^2 \left\{ 1-\frac{s_1+1}{c_1+1} \right\} \right]. \label{Fnoenv2}
\end{eqnarray}
(\ref{Hom0}), (74), (\ref{Fnoenv2}) can also be directly obtained from (63) by taking the limit of
$r_2 \rightarrow \infty$.
 The same expression of $F_{av}$ as in (65) follows, consistent with the previous result for a generic initial and final squeezed state.
In particular, in the absence of the environment, we achieve
the unit fidelity for the quantum teleportation with the initial and final EPR state, which is the case originally studied in \cite{Vaidman94}.
We now consider the effect of the environment in this special case.
Bob's state after measurement can be written similarly as in (\ref{BobDM}) as
\begin{eqnarray}
 \hat{\rho}_{B} = \frac{1}{\pi} \sum_{n,m=0}^{\infty}~_A\langle n |_C\langle n | \hat{D}^{\dagger}(\beta) \hat{\rho}_{ABC}  \hat{D}(\beta)|m \rangle_A|m \rangle_C.
\label{BobABC}
\end{eqnarray}
Here we consider the influence of environment on all three observer's states and wrote their state in the density matrix form as $\hat{\rho}_{ABC}$.
We obtain the averaged fidelity as
\begin{eqnarray}
F_{av}&=& \frac{1}{\left[ (\tilde{\Sigma}_{11}+1)(\tilde{\Sigma}_{22}+1)
 -\tilde{\Sigma}_{12}^2 \right]^{1/2}},  \label{FavEPR}
\end{eqnarray}
where $\tilde{\Sigma}_{ij}$ are components of a $2\times 2$ matrix
made up of the components of the $6\times 6$ matrix ${\bf \Large \Sigma}$ in (21).
We first write ${\bf \Large \Sigma}$ as
\begin{eqnarray}
\left( \begin{array}{ccc}
      \Sigma_{B} & 0&0\\
      0 & \Sigma_{A}&\Sigma_{AC}\\
      0 & \Sigma_{AC}^T&\Sigma_{C}
            \end{array}      \right)
=\left( \begin{array}{cccccc}
      \Sigma_{11} & \Sigma_{12}
     &0&0&0&0\\
       \Sigma_{21}&\Sigma_{22}
     &0&0&0&0\\
       0&0&\Sigma_{33}&\Sigma_{34}&\Sigma_{35}&\Sigma_{36}\\
       0&0&\Sigma_{43}&\Sigma_{44}&\Sigma_{45}&\Sigma_{46}\\
       0&0&\Sigma_{53}&\Sigma_{54}&\Sigma_{55}&\Sigma_{56}\\
       0&0&\Sigma_{63}&\Sigma_{64}&\Sigma_{65}&\Sigma_{66}
            \end{array}      \right).
   \label{SigmaEPR}
\end{eqnarray}
Then
\begin{eqnarray}
\tilde{\Sigma}&=& \sigma_3 \Sigma_{B} \sigma_3+\Sigma_{A}
-\Sigma_{AC}\sigma_3-\sigma_3\Sigma_{AC}^T + \sigma_3 \Sigma_{C} \sigma_3. \label{tildeSigma}
\end{eqnarray}
\begin{figure}[h]
 \begin{center}
\epsfxsize=0.5\textwidth \epsfbox{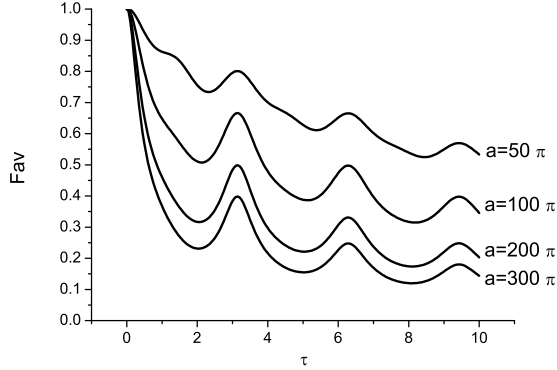}
\end{center}
\caption{
The temporal evolution of the fidelity for the maximally entangled initial and final state is shown. $\gamma=0.001$,~$\Lambda=50$.
\label{fig12}}
\end{figure}
In Fig. 12, the time evolution of the average fidelity for the
maximally entangled initial and final state is plotted. The decay
is faster for larger acceleration. After the damped oscillatory
behavior, all curves appear to reach nonvanishing asymptotic
values. We will now show this is indeed the case.

In the long time limit, the state of Bob in the absence of Alice and Chris will thermalize at the Unruh temperature $T_U$ similarly to the single particle quantum brownian motion at finite temperatures\cite{QBM0}.
On the other hand, the state of Alice and Chris in the absence of Bob will be in the steady oscillatory state (see the discussion in \cite{Shiokawa08}).
Now let us first assume the interaction of Bob with the environment is negligible.
In this case, $\Sigma_{B}=0$. In the weak coupling limit\cite{Shiokawa08},
\begin{eqnarray}
\Sigma_{A}=\Sigma_{AC}=\Sigma_{C}=\frac{1}{4}\left( \begin{array}{cc}
      1 & 0\\
      0 & 1 \end{array}      \right).\label{SigmaNoB}
\end{eqnarray}
Then
\begin{eqnarray}
\tilde{\Sigma}&=&\left( \begin{array}{cc}
      0 & 0\\
      0 & 1 \end{array}\right)
       \label{tildeSigmaEPR}
\end{eqnarray}
follows.
This will give the averaged fidelity
\begin{eqnarray}
F_{av}&\rightarrow& \frac{1}{\sqrt{2}} \sim 0.70.\label{FavEPRnoB}
\end{eqnarray}
Thus although there is still information loss, a part of quantum
information manages to escape making quantum teleportation
successful\cite{Furusawa:98}. This case can be compared to
\cite{GottesmanPreskill04}, where the authors pointed out that the
unitary quantum gates between the original qubit with one of the
Bell pair prior to the measurement ruin the unitarity of the
transfer matrix, and to \cite{Lloyd06}, where the author included
random unitary gates in quantum circuits estimated the fidelity to
be $0.85$.

Now we take into account the interaction between Bob and the environment.
This will add the thermal factor in the induced fluctuations of Bob's state in the long time limit as
\begin{eqnarray}
\Sigma_{B}=\frac{\coth(\Omega/2T_U)}{4}\left( \begin{array}{cc}
      1 & 0\\
      0 & 1 \end{array}      \right). \label{SigmaBABC}
\end{eqnarray}
Then
\begin{eqnarray}
\tilde{\Sigma}&=&\frac{\coth(\Omega/2T_U)}{4}\left( \begin{array}{cc}
      1 & 0\\
      0 & 1 \end{array}      \right)+
      \left( \begin{array}{cc}
      0 & 0\\
      0 & 1 \end{array}\right).
       \label{tildeSigmaEPR2}
\end{eqnarray}
The averaged fidelity in this case will decrease to
\begin{eqnarray}
F_{av}=\frac{2}{(\coth(\Omega/2T_U)+4)^{1/2}(\coth(\Omega/2T_U)+8)^{1/2}} \leq \frac{2}{\sqrt{45}}\sim 0.30 < 0.5, \label{FavABCfinal}
\end{eqnarray}
where the equality holds for zero acceleration ($T_U=0$).
  We see that quantum teleportation fails in this case. In this sense, quantum information will be totally lost.

\section{Conclusion} \label{Conclusion}

We studied the quantum entanglement dynamics of
static and accelerated observers. Due to the interaction with the environment,
quantum states of the observer suffer decoherence causing their entanglement to decay.
In addition, due to accelerated motion, the observer sees the vacuum as a thermal state, which enhances the rate of disentanglement in observers' quantum states moving in the vacuum.

We derived uncertainties for the quantum states of two and three observers
and saw their dynamical evolution.
Since our calculations are exact, uncertainty relations
are always satisfied as they should be. They are no longer guaranteed under perturbative approximations.
The evolution of uncertainties shows damped oscillations whose amplitudes are enhanced by the Unruh effect.
The larger the acceleration of the observer, the larger the oscillation of uncertainties because the effective temperature (Unruh temperature) for the thermal fluctuations is proportional to the acceleration.

The entanglement dynamics between two observers,
one static and the other under constant acceleration was analyzed.
From the uncertainty of the partially transposed density matrix,
the negativity and the log-negativity were calculated.
They showed that the Unruh effect on Bob dynamically enhances the uncertainty
after the partial transpose leading to the suppression of bipartite entanglement
between Alice and Bob' state.

We further studied tripartite entanglement between two static and
one accelerated observer. The acceleration modifies the three-mode
uncertainty similarly to the two-mode uncertainty. The three-mode
uncertainty shows the damped oscillations while satisfying the
uncertainty relation.
 The environment dynamically modifies initial entanglement between Alice and Bob
 and induces entanglement between Alice and Chris.
 As a result, bipartite entanglement $E_{AB}$ and $E_{AC}$ change in time
 in an opposite manner. No bipartite entanglement is induced between Bob and Chris. Bipartite entanglement is suppressed by acceleration of observers.
From the combinations of bipartite entanglement among three
parties, the genuine tripartite entanglement $E_{ABC}$ was
calculated. The positiveness of $E_{ABC}$ indicates that the
monogamy inequality is always satisfied. For our initial
condition, a product state of a two-mode squeezed state of Alice
and Bob and a coherent state of Chris, $E_{ABC}$ vanishes
initially. It becomes positive due to the effective interaction
between Alice and Chris induced from the vacuum at intermediate
time scale. In a meantime, decoherence induced from the
environment causes $E_{ABC}$ to decay at long times. Similarly to
the bipartite entanglement dynamics\cite{Shiokawa08}, the
environment plays a dual role. It induces multipartite
entanglement among subsystems but it also is the source of
decoherence and dissipation that decrease entanglement causing
information loss.

Next we studied quantum teleportation protocols between two static
and one accelerated observer. We first looked at the case with
arbitrary final states of Alice and Bob by summing over all
possible final states of them. What we obtained is an one-mode Q
function for Bob's state. In the long time and weak coupling limit
for a zero acceleration, we saw that the fidelity reaches one and
there is no information loss in this case. Then we considered the
generalized final state to which the quantum state of Alice and
Chris projected after the measurement to be an arbitrary two-mode
squeezed state with the squeezing parameter $r_2$. The
time-dependent teleportation fidelity is calculated for the
measure of success of quantum teleportation. We showed that after
performing the proper unitary transformation on Bob's quantum
state based on the measurement result sent to Bob, the fidelity is
recovered to a unit value in the limit of $r_1, r_2\rightarrow
\infty$ in the absence of the environment. For the generic value
of $r_1, r_2$, however, the fidelity is always smaller than $1$.
Thus there is information loss even in the absence of the
interaction. Furthermore, even if the initial state of Alice and
Bob is maximally entangled, the fidelity is still smaller than $1$
for generic finite values of $r_2$.

In the presence of the interaction with the vacuum, the fidelity
shows damped oscillations similarly to entanglement. This
indicates the close relation between entanglement and fidelity. In
the absence of environment, there is a direct relation between
fidelity and entanglement for maximally entangled final states as
seen in (\ref{FavPS}). With the generic final state in the
presence of the environment, the relation is not so
straightforward.
 The damping of fidelity is enhanced by the acceleration.
 The Unruh effect overall suppresses fidelity causing more information loss.
We obtained the fidelity expression for the maximally entangled
final state case $r_2\rightarrow \infty$ in (\ref{Favmaxfinal}).
This type of final states is usually assumed in current
experiments for quantum teleportation\cite{HIKF01}. We saw that
the temporal behavior of the fidelity consists of damped
oscillations  similarly to the case with generic $r_2$. The
maximum value of the fidelity is achieved only at the initial
time, which becomes a unit value as $r_1$ goes to infinity.


We also examined the case when the initial and the final state are
both maximally entangled ($r_1, r_2\rightarrow \infty$). In this
case, the teleportation is perfect (the fidelity reaches the unit
value) in the absence of the environment or other sources of
information loss. In the presence of the interaction with vacuum,
however, the fidelity is shown to decay. Taking the long time
limit, we obtained the asymptotic value of the fidelity. If we
ignore the interaction of Bob's state with the vacuum, we obtain
$F_{av}\sim 0.7$ and quantum teleportation can still be viewed as
successful. Proper encoding of the message with quantum error
correction may help retrieving the complete information.
 If we take into account the interaction with vacuum for all three
parties, we obtain $F_{av}\sim 0.3$, thus quantum teleportation is
no longer successful.

Implications of our results to black hole information problem can
be summarized as follows. There are two fundamental sources of
nonunitary evolutions and information loss. (1) the boundary
condition of the evolution of the total system (intrinsic loss).
(2) the interaction of the relevant quantum mechanical system with
the surrounding quantum field (induced loss). (1) can be
attributed in the laboratory to the special choice of boundary
conditions imposed by measurement. (2) is present in any spacetime
and responsible for the quantum/classical correspondence of
quantum systems of large degrees of freedom\cite{Dec96}. In a
black hole spacetime, the singularity provides an effective
boundary on the quantum state of infalling matter and the incoming
Hawking particle. The origin of nonunitary evolution in (1) may be
explained by the remnants or wormholes\cite{Page93}. (2) is also
present as the interaction among infalling matter, Hawking
particles, and gravitational field of a black hole. Thus in
addition to the correlation in a Hawking pair, there is the
induced three-body correlation and tripartite entanglement between
the matter and a Hawking pair, which vary dynamically due to
gravitational interaction. In the practical setting of the
laboratory, there are many other phenomenological loss of
information such as photon loss during the optical transmission,
leakage, etc. We did not consider these effects in this work.

At the present stage without complete knowledge of quantum
gravity, we do not know details about the final state of the black
hole. Assuming a maximally entangled final state, we may retrieve
quantum information from a black hole just as in quantum
teleportation as suggested in \cite{MaldaneceHorowitz04} provided
that (I) the induced loss due to vacuum is negligible. Since the
environment is always present, this type of loss is unavoidable.
(II) The proper postselection is performed.  This can be viewed as
an extension of the time-neutral formulation of quantum
mechanics\cite{ABL64} applied to spacetime physics, in which both
initial and final states are specified. Decoherence
history interpretation of the formulation in view of cosmology is given in
\cite{GellMannHartle93}. In the meantime, performing the
postselection in the laboratory is now very common in quantum
information science. We hope our attempts using accelerated
observers provide a practical setting to understand the black hole
information problem.


\end{document}